\documentclass[aps,prc,groupedaddress,twocolumn,nofootinbib]{revtex4-2}

\usepackage{bm}
\usepackage{graphicx}
\usepackage{ascmac}
\usepackage{amsmath,amssymb}
\usepackage{cancel}
\usepackage{braket}

\begin{document}


\title{Structure of exotic hadrons by a weak-binding relation with finite-range correction}


\author{Tomona Kinugawa}
\email[]{kinugawa-tomona@ed.tmu.ac.jp}
\affiliation{Department of Physics, Tokyo Metropolitan University, Hachioji 192-0397, Japan}
\author{Tetsuo Hyodo}
\email[]{hyodo@tmu.ac.jp}
\affiliation{Department of Physics, Tokyo Metropolitan University, Hachioji 192-0397, Japan}

\date{\today}

\begin{abstract}
The composite nature of a shallow bound state is studied by using the weak-binding relation, which connects the compositeness of the bound state with observables. We first show that the previous weak-binding relation cannot be applied to the system with a large effective range. To overcome this difficulty, we introduce the finite-range correction by redefining the typical length scale in the weak-binding relation. A method to estimate the uncertainty of the compositeness is proposed. It is numerically demonstrated that the range correction enlarges the applicable region of the weak-binding relation. Finally, we apply the improved weak-binding relation to the actual hadrons, nuclei, and atomic systems [deuteron, $X(3872)$, $D^{*}_{s0}(2317)$, $D_{s1}(2460)$, $N\Omega$ dibaryon, $\Omega\Omega$ dibaryon, ${}^{3}_{\Lambda}{\rm H}$, and ${}^{4}{\rm He}$ dimer] to discuss their internal structure from the compositeness. We present a reasonable estimation of the compositeness of the deuteron by properly taking into account the uncertainty. The results of $X(3872)$ and the $N\Omega$ dibaryon show that the range correction is important to estimate the compositeness of physical states.

\end{abstract}


\maketitle


\section{Introduction}
\label{sec:intro}

Constituent quark models describe hadrons as the $qqq$ or $q\bar{q}$ states, and they have succeeded in explaining the spectrum of most of observed hadrons~\cite{Isgur:1978xj,Godfrey:1985xj}. However, recent experiments imply that some states in the heavy quark sector have an exotic internal structure, other than $qqq$ or $q\bar{q}$~\cite{Hosaka:2016pey,Guo:2017jvc,Brambilla:2019esw}. It is considered that those states are, for instance, the multiquark states which are the compact states with at least four quarks, hadronic molecular states which are the weakly bound states of hadrons, or gluon hybrid states which contain the constituent gluons. The intensive studies with both the theoretical and experimental approaches have been performed to clarify the nature of the candidates for exotic hadrons. 

One of the most actively studied candidates for exotic hadrons is $X(3872)$, observed in the invariant mass spectrum of $\pi^{+}\pi^{-}J/\psi$ in the $B^{\pm}\to K^{\pm}\pi^{+}\pi^{-}J/\psi$ decay in 2003 by the Belle Collaboration~\cite{Belle:2003nnu}. Because $X(3872)$ was observed in the lower energy region than $\chi_{c1}'$ predicted by the quark model, it is expected to have an exotic structure. In addition to $X(3872)$, many states have been observed in the higher energy regions than the $D\bar{D}$ threshold. These states do not fit well in the prediction by the quark model, and are called $XYZ$ mesons~\cite{Brambilla:2019esw}.

In 2021, $T_{cc}$ was observed in the invariant mass spectrum of $D^{0}D^{0}\pi^{+}$ in the $pp$ collision by the LHCb Collaboration~\cite{LHCb:2021vvq,LHCb:2021auc}. Because $T_{cc}$ has the charm $C=+2$ and the minimal quark content is $cc\bar{u}\bar{d}$, $T_{cc}$ is an exotic state with at least four quarks. 

To study the internal structure of the candidates for exotic hadrons theoretically, we employ the weak-binding relation~\cite{Weinberg:1965zz,Kamiya:2015aea,Kamiya:2016oao} as a model-independent approach. For this purpose, we introduce the compositeness $X$~\cite{Hyodo:2013nka} which is the quantitative weight of the hadronic molecular component in the physical hadron wavefunction. Denoting the hadronic molecular component schematically as $\ket{\rm molecule}$, we can express $X$ as the probability to find the hadronic molecular component in the normalized wavefunction of the bound state $\ket{\Psi}$:
\begin{align}
X&=|\braket{{\rm molecule}|\Psi}|^{2}.
\end{align}
It was shown in Refs.~\cite{Weinberg:1965zz,Kamiya:2015aea,Kamiya:2016oao} that $X$ and the observables satisfy the weak-binding relation:
\begin{align}
a_0=R\left\{\frac{2X}{1+X}+\mathcal{O}\left(\frac{R_{\rm typ}}{R}\right)\right\}.
\label{eq:wbr_s}
\end{align}
 Here $a_{0}$ is the scattering length, $R\equiv1/\sqrt{2\mu B}$ is the radius of the bound state, determined by the binding energy $B$ and the reduced mass $\mu$, and $R_{\rm typ}$ is the typical length scale of the interaction, for instance, the range of the interaction. When $B$ is sufficiently small and $R\gg R_{\rm typ}$, we can neglect the correction terms $\mathcal{O}({R_{\rm typ}}/{R})$ in Eq.~\eqref{eq:wbr_s}. Then, the weak-binding relation reduces to $a_0=R \{2X/(1+X)\}$, and $X$ is model-independently estimated only by the observables $a_{0}$ and $B$ for the weakly bound states. 

In general, the low-energy universality holds~\cite{Braaten:2004rn,Naidon:2016dpf} when the scattering length $a_{0}$ is sufficiently larger than other length scales. If a system has a large scattering length,  the microscopic details of the system can be neglected, and the different systems follow the same universal laws. For example, for two-body systems with a shallow bound state, the low-energy universality implies the relation $a_{0}=R$. For three-body systems, the Efimov effect~\cite{Efimov:1971zz,Braaten:2004rn,Naidon:2016dpf} is known. In practice, when the effective range $r_{e}$ is not negligible in comparison with $a_{0}$, there arises the deviation from the consequence of the universality, called the range correction.

In this paper, we discuss the range correction in the weak-binding relation~\eqref{eq:wbr_s}.
We first show that Eq.~\eqref{eq:wbr_s} does not hold when the effective range $r_{e}$ is large which indicates the necessity of the range correction.  We then propose the improvement of the weak-binding relation with the range correction so that it becomes appricable to the system with a large $r_{e}$. Finally, we discuss the internal structure of the actual physical states using the improved weak-binding relation with the range correction.

The compositeness and the weak-binding relation were proposed by Weinberg to analyze the nature of the deuteron~\cite{Weinberg:1965zz}, and have been applied to many hadron systems since around 2000. The model-independent estimations of the compositeness from the weak-binding relation are proposed in Refs.~\cite{Baru:2003qq,Hanhart:2011jz,Hyodo:2013iga,Kamiya:2015aea,Kamiya:2016oao,Matuschek:2020gqe}, and the calculations of the compositeness from the residue of the poles are also performed in Refs.~\cite{Hyodo:2011qc,Aceti:2012dd,Hanhart:2014ssa,Sekihara:2014kya,Guo:2015daa,Sekihara:2015gvw,Sekihara:2016xnq}. 
Some recent studies discuss the range corrections in the weak-binding relation focusing on the compositeness of the deuteron. In Ref.~\cite{Li:2021cue}, the authors introduce a general form factor of the bound state and discuss the range correction through the form factor.
The model with a finite cutoff is considered in Ref.~\cite{Song:2022yvz}. By comparing the effective range in the model with the experimental value to determine the value of the cutoff, they estimate the compositeness.
In Ref.~\cite{Albaladejo:2022sux}, the authors introduce the effects of $R_{\rm typ}$ to $a_{0}$ and $r_{e}$ as the phenomenological term, and estimate the likelihood of the compositeness model-independently.

This paper is organized as follows. For later discussions, in Sec.~\ref{sec:eft}, we introduce the nonrelativistic effective field theory which has the derivative-coupling interaction and the coupling with the discrete state. In Sec.~\ref{sec:rangecorrection}, we discuss the weak-binding relation in the system with a non-negligible effective range, and propose the range correction by modifying the correction terms. We perform the numerical calculations to check the validity of the range correction in Sec.~\ref{sec:numerical}. In Sec.~\ref{sec:apply}, the improved weak-binding relation is applied to the actual hadrons, hypernuclei, and atomic systems to discuss their internal structure by focusing on the significance of the range correction. Section~\ref{sec:sum} is devoted to summary.  The preliminary results of Secs.~\ref{sec:rangecorrection}, \ref{sec:numerical}, and \ref{sec:apply} are partly reported in Refs.~\cite{Kinugawa:2021ykv}, \cite{Kinugawa:2021ybb}, and \cite{Kinugawa:2022ohs}, respectively.

\section{Effective field theory}
\label{sec:eft}
We introduce a nonrelativistic effective field theory to discuss the low-energy limit of two-hadron-scattering systems with a weakly bound state in this section. In general, effective field theories describe the low-energy phenomena of some microscopic theories~\cite{Braaten:2007nq}. Here, we consider the scattering of two identical bosons with the following Hamiltonian:
\begin{align}
\mathcal{H}&=\frac{1}{2m}{\nabla}\psi^{\dag}\cdot {\nabla}\psi+\frac{1}{4m}{\nabla}\phi^{\dag}\cdot {\nabla}\phi+\nu_{0}\phi^{\dag}\phi \nonumber \\
&\quad +\frac{1}{4}\lambda_{0}(\psi^{\dagger}\psi)^{2}+\frac{1}{4}\rho_{0}\nabla(\psi^{\dagger}\psi)\cdot \nabla(\psi^{\dagger}\psi) \nonumber \\
&\quad +\frac{1}{2}g_{0}(\phi^{\dagger}\psi^{2}+\psi^{\dagger 2}\phi),
\label{eq:H_effectiveresonance}
\end{align}
where $\psi$ is the boson field, $\phi$ is the discrete state coupled with the two-boson scattering, $m$ is the boson mass, and $\nu_{0}$ is the energy of $\phi$ measured from the two-$\psi$ threshold. $\lambda_{0}$ and $\rho_{0}$ are the bare coupling constants of the two-$\psi$ contact interaction and the derivative coupling term, respectively. $g_{0}$ is the coupling constant of the two-body $\psi$ scattering and the discrete state $\phi$. We introduce the momentum cutoff $\Lambda$ to remove the divergence of the momentum integration of the intermediate bosons in the Lippmann-Schwinger equation, which stems from the contact interactions in Hamiltonian~\eqref{eq:H_effectiveresonance}. This theory is renormalizable; we can take the limit $\Lambda\to \infty$ with keeping the observables finite. This corresponds to the zero-range limit, because the inverse of the cutoff is regarded as the interaction range. Here, we leave $\Lambda$ finite so that the interaction range is also finite.

Because the interaction terms of the Hamiltonian~\eqref{eq:H_effectiveresonance} have no angular dependence, the scattering occurs only in the $s$ wave. The two-$\psi$ scattering amplitude $f(k)$ at momentum $k$ is obtained as
\begin{widetext}
\begin{align}
&f(k)=\left[-\frac{8\pi}{m}\frac{(1+\frac{m}{12\pi^{2}}\rho_{0}\Lambda^{3})^{2}}{(\lambda_0+\frac{g_0^2}{E-\nu_0}-\frac{m}{20\pi^{2}}\rho_{0}^{2}\Lambda^{5})+2\rho_{0}(1+\frac{m}{24\pi^{2}}\rho_{0}\Lambda^{3})k^{2}}-\frac{2\Lambda}{\pi}-ik\right]^{-1},
\label{eq:f_effectiveresonance}
\end{align}
with $E=k^{2}/m$. Here we neglect the terms which vanish in the $\Lambda\to \infty$ limit in the momentum integration. The scattering length $a_{0}$ and the effective range $r_{e}$ are read off by comparing it with the effective range expansion:
\begin{align}
f(k)&=\left[-\frac{1}{a_{0}}+\frac{r_{e}}{2}k^{2}+\cdots-ik\right]^{-1}.
\label{eq:ere}
\end{align}
Then we obtain 
\begin{align}
a_{0}&=\left[\frac{8\pi}{m}\frac{(1+\frac{m}{12\pi^2}\rho_0\Lambda^3)^2}{\lambda_0-\frac{m}{20\pi^2}\rho_0^2\Lambda^5-\frac{g_0^2}{\nu_0}}+\frac{2}{\pi}\Lambda\right]^{-1},
\label{eq:a0_effectiveresonance}\\
r_{e}&=\frac{16\pi}{m}\frac{(1+\frac{m}{12\pi^2}\rho_0\Lambda^3)^2\left[2\rho_0(1+\frac{m}{24\pi^2}\rho_0\Lambda^3)-\frac{g_0^2}{m\nu_0^{2}}\right]}{\left(\lambda_0-\frac{m}{20\pi^2}\rho_0^2\Lambda^5-\frac{g_0^2}{\nu_0}\right)^2}.
\label{eq:re_effectiveresonance}
\end{align}
\end{widetext}
When there is a bound state in the two-body scattering, the scattering amplitude $f(k)$ has a pole at $k=k_{B}$ with a pure imaginary $k_{B}$. The radius of the bound state $R$ is then given as $R=i/k_{B}$.

We note that this model reduces to the effective range model, the resonance model, and the zero-range model introduced in Ref.~\cite{Braaten:2007nq}, by adjusting the coupling constants $g_{0}$ and $\rho_{0}$ appropriately. With $g_{0}=0$ and $\rho_{0}\neq0$, we obtain the effective range model. The scattering amplitude in this model is 
\begin{align}
f(k)&=\left[-\frac{8\pi}{m}\frac{(1+\frac{m}{12\pi^{2}}\rho_{0}\Lambda^{3})^{2}}{(\lambda_0-\frac{m}{20\pi^{2}}\rho_{0}^{2}\Lambda^{5})+2\rho_{0}(1+\frac{m}{24\pi^{2}}\rho_{0}\Lambda^{3})k^{2}}\right. \nonumber \\ 
&\quad \left.-\frac{2\Lambda}{\pi}-ik\right]^{-1}.
\label{eq:f_effective}
\end{align}
From Eq.~\eqref{eq:f_effective}, $a_{0}$ and $r_{e}$ are 
\begin{align}
a_{0}&=\left[\frac{8\pi}{m}\frac{(1+\frac{m}{12\pi^2}\rho_0\Lambda^3)^2}{\lambda_0-\frac{m}{20\pi^2}\rho_0^2\Lambda^5}+\frac{2}{\pi}\Lambda\right]^{-1},
\label{eq:a0-rho-lambda}\\
r_{e}&=\frac{16\pi}{m}\frac{(1+\frac{m}{12\pi^2}\rho_0\Lambda^3)^2\left[2\rho_0(1+\frac{m}{24\pi^2}\rho_0\Lambda^3)\right]}{\left(\lambda_0-\frac{m}{20\pi^2}\rho_0^2\Lambda^5\right)^2}.
\label{eq:re-rho-lambda}
\end{align}
In this model, the effective range $r_{e}$ can take any negative value. Positive $r_{e}$ is, however, allowed only below an upper bound (Wigner bound) as discussed in Sec.~ \ref{sec:er_model_stable}. With a finite cutoff $\Lambda$, the scattering amplitude contains higher order terms of $\mathcal{O}(k^{4})$. By taking the zero-range limit ($\Lambda\to \infty$) after the renormalization, these terms vanish and the scattering  amplitude is expressed only by $a_{0}$ and $r_{e}$.

With $g_{0}\neq0$ and $\rho_{0}=0$, the resonance model is obtained where two identical bosons scatter with a coupling to the discrete state. This model gives
\begin{align}
a_{0}&=\left[\frac{8\pi}{m}\left(\lambda_{0}-\frac{g_{0}^{2}}{\nu_{0}}\right)^{-1}+\frac{2}{\pi}\Lambda\right]^{-1},
\label{eq:a0-g-mu}\\
r_{e}&=-\frac{16\pi g_{0}^{2}}{m^{2}\nu_{0}^{2}}\left(\lambda_{0}-\frac{g_{0}^{2}}{\nu_{0}}\right)^{-2}.
\label{eq:re-g-mu}
\end{align}
In contrast to the effective range model, $r_{e}$ is always negative in this model, as seen in Eq.~\eqref{eq:re-g-mu}.

Furthermore, with $g_0=0$ and $\rho_{0}=0$, in other words, with no effects of the derivative couplings and the discrete states, we reproduce the zero-range model whose scattering amplitude is expressed only by $a_{0}$. In this case, Eqs.~\eqref{eq:a0_effectiveresonance} and \eqref{eq:re_effectiveresonance} reduce to 
\begin{align}
a_{0}&=\frac{m}{8\pi}\frac{1}{1/\lambda_0+(m/4\pi^2)\Lambda},
\label{eq:a-zerorange}\\
r_{e}&=0.
\end{align}
We utilize those three models for the analytical and numerical calculations in later sections.

\section{Range correction}
\label{sec:rangecorrection}
We discuss the range correction in the weak-binding relation in this section. In Sec.~\ref{sec:rangecorrectionsdiscussiton}, we first give the physical interpretation of each term in the weak-binding relation~\eqref{eq:wbr_s} from the viewpoint of the low-energy universality. We then classify the property of the exact values of the compositeness of the models introduced in the previous section in Sec.~\ref{sec:models-X}. In Sec.~\ref{sec:discussion-rangecorrection}, we show that the contradiction arises when we apply the weak-binding relation to the effective range model in the zero-range limit. To solve the above problem, we propose the range correction to the weak-binding relation in Sec.~\ref{subsec:improve}.
A practical method to estimate the uncertainty of the compositeness is presented in Sec.~\ref{subsec:uncertainty}.

\subsection{Low-energy universality and weak-binding relation}
\label{sec:rangecorrectionsdiscussiton}

Let us consider the weak-binding relation from the perspective of the low-energy universality. It is known that the radius of the bound state $R$ is scaled by the scattering length $a_{0}$ as a consequence of the low-energy universality~\cite{Braaten:2004rn,Naidon:2016dpf}:
\begin{align}
R&\equiv\frac{1}{\sqrt{2\mu B}}=a_{0}.
\label{eq:R=a0}
\end{align}
Both the scattering length $a_{0}$ and the radius $R$ diverge when $B\to 0$ as shown in Eq.~\eqref{eq:R=a0}. In this weak-binding limit, the weight of the two-body composite state dominates the whole wavefunction ($X\to1$)~\cite{Hyodo:2014bda}. This is consistent with the consequence of the weak-binding relation with $X=1$ and $a_{0}=R$ for $R\to \infty$. 

When the binding energy is finite and the system is away from the weak-binding limit ($R<\infty$), the deviations from Eq.~\eqref{eq:R=a0} arise. The origins of the deviations are induced by two terms in Eq.~\eqref{eq:wbr_s}: (i) the $2X/(1+X)$ term, with the deviation due to the contributions from other channels ($X<1$), and (ii) the $\mathcal{O}(R_{\rm typ}/R)$ term, with the deviation due to the finite interaction range ($R_{\rm typ}\neq 0$). Therefore, when either one of those contributions exists, $a_{0}\neq R$. Now we consider the effects of those contributions separately. The contributions from other channels make $X$ smaller than unity by the definition (see Sec.~\ref{sec:models-X}). Hence, with the zero-range interaction ($R_{\rm typ}=0$), $R$ is always larger than $a_{0}$ because $2X/(1+X)<1$. On the other hand, in the single-channel case ($X=1$), $R$ can be not only larger than but also smaller than $a_{0}$ because the sign of the correction terms $\mathcal{O}(R_{\rm typ}/R)$ is arbitrary. Therefore, for a system with $R<a_{0}$, the effects of the finite interaction range should play an important role. In the case of the deuteron in the $NN$ scattering, actually, $R=4.31$ fm and $a_{0}=5.42$ fm (i.e., $a_{0}>R$), suggesting the importance of the finite interaction range~\cite{Li:2021cue,Song:2022yvz,Albaladejo:2022sux}. 

\subsection{Compositeness in effective field theories}
\label{sec:models-X}

We can calculate the exact values of the compositeness $X_{\rm exact}$ in the models introduced in Sec.~\ref{sec:eft}. Let us consider $X_{\rm exact}$ of the zero-range model, effective range model, and resonance model to compare them with the compositeness $X$ estimated by the weak-binding relation in the later discussions. In the zero-range model and the effective range model which have no couplings with other channels, the completeness relation is given by
\begin{align}
\int \frac{d^{3}{k}}{(2\pi)^{3}}\ket{\bm{k}}\bra{\bm{k}}=1,
\label{eq:complete-zrm-erm}
\end{align}
where $\ket{\bm{k}}$ is the eigenstate of the free Hamiltonian with the relative momentum $\bm{k}$ in the $\psi\psi$ scattering. However, the completeness relation of the resonance model which has the couplings with the discrete state $\ket{\phi}$ is as follows:
\begin{align}
\int \frac{d^{3}{k}}{(2\pi)^{3}}\ket{\bm{k}}\bra{\bm{k}}+\ket{\phi}\bra{\phi}=1,
\label{eq:complete-rm}
\end{align}
because the eigenstates of the free Hamiltonian consist of the discrete state $\ket{\phi}$ in addition to $\ket{\bm{k}}$.

The definition of the compositeness $X$ is given by~\cite{Kamiya:2016oao}
\begin{align}
X&=\int\frac{d^{3}k}{(2\pi)^{3}}|\braket{\bm{k}|\Psi}|^{2},
\label{eq:def-X}
\end{align}
where the bound state $\ket{\Psi}$ is the eigenstate of the full Hamiltonian. 
From the completeness relation~\eqref{eq:complete-zrm-erm} and the normalization of the bound state $\braket{\Psi|\Psi}=1$, definition~\eqref{eq:def-X} gives $X_{\rm exact}$ of the zero-range model and effective range model:
\begin{align}
X_{\rm exact}&=1.
\end{align}
With the same reasoning, $X_{\rm exact}=1$ in any single-channel models. In contrast, $X_{\rm exact}$ of the resonance model is
\begin{align}
X_{\rm exact}&<1,
\end{align}
because $\braket{\Psi|\phi}\braket{\phi|\Psi}>0$ unless $g_{0}=0$. In fact, we will derive the explicit expression in Eq.~\eqref{eq:rm-X-exact-2} which shows $X_{\rm exact}<1$ except for a special case with $\nu_{0}=0$.

For later convenience, we also summarize the properties of the effective range $r_{e}$ in each model. As mentioned in Sec.~\ref{sec:eft}, $r_{e}=0$ in the zero-range model, $r_{e}\neq0$ in the effective range model, and $r_{e}<0$ in the resonance model. We tabulate $r_{e}$ and $X_{\rm exact}$ of each model in Table~\ref{tab:re-X}. We note that while both the effective range model and the resonance model have $r_{e}\neq0$, $X=1$ in the effective range model and $X\neq1$ in the resonance model.

 \begin{table}
 \caption{The effective range $r_{e}$ and the exact value of the compositeness $X_{\rm exact}$ in the zero-range model, effective range model, and resonance model.\label{tab:re-X}}
 \begin{ruledtabular}
 \begin{tabular}{lcc}
Effective field theory & $r_{e}$ & $X_{\rm exact}$ \\ \hline
Zero-range model & $r_{e}=0$ & $X=1$ \\ 
Effective range model & $r_{e}\neq0$ & $X=1$ \\ 
Resonance model & $r_{e}<0$ & $X<1$ \\ 
 \end{tabular}
 \end{ruledtabular}
 \end{table}

\subsection{Deviation from low-energy universality by effective range}
\label{sec:discussion-rangecorrection}

We now consider the role of the effective range in the weak-binding relation. The weak-binding relation~\eqref{eq:wbr_s} implies $a_{0}=R$, in accordance with Eq.~\eqref{eq:R=a0}, for any single-channel models ($X=1$) with the zero-range interaction ($R_{\rm typ}= 0$). However, we show that the effective range model does not lead to $a_{0}=R$ in the zero-range limit. To see this, we derive the scattering amplitude in the zero-range limit by taking the $\Lambda\to \infty$ limit in the renormalized scattering amplitude of the effective range model:
\begin{align}
f(k)=\left[-\frac{1}{a_0}+\frac{r_e}{2}k^2-ik\right]^{-1}.
\label{eq:f-1-zero}
\end{align}
To relate $a_{0}$ and $R$ in this model, we search for the pole of $f(k)$. Because the bound-state pole appears at $k=i/R$, we express $a_{0}$ by $R$ and $r_{e}$ using the bound-state condition
\begin{align}
a_{0}&=R\frac{1}{1-r_e/(2R)}
\label{eq:relations}\\
&=R\left[1+\mathcal{O}\left(\frac{|r_e|}{R}\right)\right],
\label{r_e_correction}
\end{align}
where we take the absolute value of $r_{e}$ because it can be negative. We find the deviation from $a_{0}=R$ by the contribution from the finite effective range $r_{e}$. This contradicts with the weak-binding relation~\eqref{eq:wbr_s} if $R_{\rm typ}$ is regarded as the interaction range. Therefore, Eq.~\eqref{r_e_correction} shows the necessity of the range correction in the weak-binding relation.

In addition, to discuss the effects of the higher order terms of the effective range expansion, we consider the zero-range model whose scattering amplitude includes the fourth order term of $k$:
\begin{eqnarray}
f(k)=\left[-\frac{1}{a_0}+\frac{r_e}{2}k^2-\frac{P_s}{4}k^4-ik\right]^{-1},
\label{eq:ere-ps}
\end{eqnarray}
where $P_{s}$ is the shape parameter with the dimension of length cubed. From the pole of  Eq.~\eqref{eq:ere-ps}, the scattering length $a_{0}$ is obtained as follows:
\begin{align}
a_0&=R\left[1+\frac{1}{2}\left(\frac{r_e}{R}\right)+\frac{1}{4}\left(\frac{r_e}{R}\right)^2+\left(\frac{r_e^3}
{8}+\frac{P_s}{4}\right)\left(\frac{1}{R}\right)^3\right.\nonumber \\
&\quad +\left.\mathcal{O}\left(\frac{|r_{e}|^{4}}{R^{4}},\frac{|r_{e}P_{s}|}{R^{4}}\right)\right].
\label{eq:a0-Ps-re}
\end{align}
In ordinary systems, the effective range $r_{e}$ and the shape parameter $P_{s}$ are estimated by the typical length scale of the interaction. For instance, we expect that $|r_{e}|\approx1~$fm and $|P_{s}|\approx 1~$fm${}^{3}$ in the case of the strong interaction. In the weak-binding case where $R$ is much larger than other scales, it is expected that ${|P_s|}/{R^3}\ll{|r_e|}/{R}$. In this case, Eq.~\eqref{eq:a0-Ps-re} reduces to Eq.~\eqref{r_e_correction} for the effective range model. 

However, if the shape parameter happens to have a large magnitude, ${|P_s|}/{R^3}$ can be much larger than ${|r_e|}/{R}$. In this case, $a_{0}$ becomes 
\begin{align}
a_{0}=R\left[1+\frac{P_s}{4}\left(\frac{1}{R}\right)^3+\mathcal{O}\left(\frac{|r_{e}|}{R}\right)\right]
=R\left[1+\mathcal{O}\left(\frac{|P_s|}{R^3}\right)\right].
\label{eq:Ps-correction}
\end{align}
This equation shows that the leading order deviation from $a_{0}=R$ can be the term of $P_{s}$. The same discussion holds when the other length scales in the effective range expansion are accidentally large. 

\subsection{Range correction in weak-binding relation}
\label{subsec:improve}
Based on the above discussion, we propose the range correction in the weak-binding relation. We show that the weak-binding relation does not reproduce Eq.~\eqref{r_e_correction} in the effective range model. To solve this problem, we consider the modification of the correction terms $\mathcal{O}(R_{\rm typ}/R)$ in Eq.~\eqref{eq:wbr_s}. 
%
%
First we define the length scale in the effective range expansion $R_{\rm eff}$ as 
\begin{align}
R_{\rm eff}&=\max\left\{|r_e|,\frac{|P_{s}|}{R^2},...\right\}.
\label{eq:Reff}
\end{align}
Then we define $R_{\rm int}$ as the interaction range which was denoted by $R_{\rm typ}$ in the previous works~\cite{Kamiya:2016oao}. As the range correction in the weak-binding relation, we propose to redefine  $R_{\rm typ}$ in the correction terms as the largest length scale among $R_{\rm int}$ and $R_{\rm eff}$:
\begin{align}
R_{\rm typ}&=\max\{R_{\rm int},R_{\rm eff}\}.
\label{eq:redef}
\end{align}
By this redefinition, the improved weak-binding relation reproduces Eq.~\eqref{r_e_correction} without any contradictions because of the nonzero correction terms with $R_{\rm eff}=|r_{e}|$. Furthermore, 
thanks to Eq.~\eqref{eq:Reff}, Eq.~\eqref{eq:Ps-correction} is also reproduced when the shape parameter $P_{s}$ is unusually large. The improved relation reduces to the previous one when $R_{\rm eff}<R_{\rm int}$. Hence, when there is a large length scale in the effective range expansion, it is necessary to use the weak-binding relation with the range correction~\eqref{eq:redef}.

Although we consider the range correction to the correction terms $\mathcal{O}(R_{\rm typ}/R)$, one may wonder about the possibility to introduce the contributions from the effective range directly to the first term in Eq.~\eqref{eq:wbr_s}. This is achieved only when the origin of $r_{e}$ can be clearly traced in the relation with the compositeness $X$. Unfortunately, this is not the case, as we show below using the model with Eq.~\eqref{eq:re_effectiveresonance}.  Table~\ref{tab:re-X} shows that there are two models with $X=1$ and $X<1$ with a finite $r_{e}$. In the effective range model, $r_{e}$ stems from the derivative couplings. Compared with the zero-range model with $X=1$, the finite $r_{e}$ arises in the effective range model without changing the value of $X$. Therefore, if $r_{e}$ comes from the derivative couplings, the value of $r_{e}$ can be taken independently of $X$. On the other hand, in the resonance model, the coupling with the discrete state induces the finite $r_{e}$. In this case, $r_{e}$ is related with $X$ because $X<1$ in the resonance model. As shown in Eq.~\eqref{eq:re_effectiveresonance} with both the derivative couplings and the coupling with the discrete state, the terms of $\rho_{0}$ and $g_{0}$ cannot be separated. Therefore, the origins of $r_{e}$ in Eq.~\eqref{eq:re_effectiveresonance} cannot be uniquely specified. In general, we cannot determine whether $X$ deviates from unity by the effective range. Thus, we need to introduce the contributions from $r_{e}$ to the correction terms.

\subsection{Uncertainty estimation}
\label{subsec:uncertainty}

The range correction~\eqref{eq:redef} is of relevance for the estimation of the uncertainty of the compositeness by the correction terms as described in Ref.~\cite{Kamiya:2016oao}. The central value of the estimated compositeness $X_{c}$ is obtained by solving Eq.~\eqref{eq:wbr_s} for $X$ by neglecting the correction terms:
\begin{align}
X_c=\frac{a_0/R}{2-a_0/R}.
\label{eq:Xc}
\end{align}
To estimate the correction terms in the weak-binding relation, we define the dimensionless quantity $\xi$ as follows~\cite{Kamiya:2016oao}:
\begin{align}
\label{eq:xi}
\xi=\frac{R_{\rm typ}}{R}.
\end{align}
To expand $a_{0}$ in powers of $\xi$ in the weak-binding relation~\eqref{eq:wbr_s}, it should be smaller than unity. Because $R_{\rm typ}$ is positive from Eq.~\eqref{eq:redef}, we also have $\xi\geq0$. Regarding the magnitude of the correction terms as $\xi$, we obtain the upper (lower) bound of the compositeness $X_{u}$ ($X_{l}$):
\begin{align}
X_u(\xi)&=\frac{a_0/R+\xi}{2-a_0/R-\xi},
\label{eq:Xu}\\
X_l(\xi)&=\frac{a_0/R-\xi}{2-a_0/R+\xi},
\label{eq:Xl}
\end{align}
for $0\leq \xi<1$. For the range correction~\eqref{eq:redef}, we need to consider both the interaction range $R_{\rm int}$ and the length scale in the effective range expansion $R_{\rm eff}$. We thus define $\xi_{\rm eff}$ and $\xi_{\rm int}$:
\begin{align}
\xi_{\rm eff}=\frac{R_{\rm eff}}{R},\quad \xi_{\rm int}=\frac{R_{\rm int}}{R}.
\end{align}
The uncertainty band of the compositeness $X$ is given by
\begin{align}
X_{l}(\xi)\leq X\leq X_{u}(\xi).
\label{eq:uncertainty-band}
\end{align}

Now we recall that the value of the compositeness $X$ of a bound state is restricted within $0\leq X\leq 1$ from the definition in Eq.~\eqref{eq:def-X}. In other words, the exact value $X_{\rm exact}$ cannot go beyond this region and therefore the uncertainty band outside the definition can be safely excluded. We thus define 
\begin{align}
\bar{X}_{u}(\xi)=\min\{X_{u}(\xi),1\},\quad \bar{X}_{l}(\xi)=\max\{X_{l}(\xi),0\} ,
\end{align}
and consider the uncertainty band consistent with the definition of the compositeness as
\begin{align}
\bar{X}_{l}(\xi)\leq X\leq  \bar{X}_{u}(\xi).
\label{eq:Xfinal}
\end{align}
We will use this uncertainty estimation in the following sections. 

\section{Numerical calculations}
\label{sec:numerical}
To check the validity of the range correction in the weak-binding relation, we perform numerical calculations with models in Sec.~\ref{sec:eft}, which have the finite interaction range $R_{\rm int}$ and the effective range $r_{e}$. For example, in the effective range model ($\rho_{0}\neq 0$ and $g_{0}=0$), we can calculate the scattering length $a_{0}$ and the effective range $r_{e}$ as shown in Eqs.~\eqref{eq:a0_effectiveresonance} and \eqref{eq:re_effectiveresonance}, and the exact value of the compositeness $X_{\rm exact}=1$ is known as discussed in Sec.~\ref{sec:models-X}. Therefore, we can study the applicability of the weak-binding relations by comparing $X_{\rm exact}$ and the compositeness $X$ estimated by the relations. 
In Sec.~\ref{sec:estimation}, through this comparison, we establish a condition for the valid application of the weak-binding relation.
We then perform the numerical calculations to discuss the applicable parameter region by using the effective range model with $X_{\rm exact}=1$ in Sec.~\ref{sec:er_model_stable}. The resonance model with $X_{\rm exact}<1$ is also examined in Sec.~\ref{sec:r-model-stable}. 
We also numerically investigate the effects of the higher order terms in the effective range expansion discussed in the previous work~\cite{Kamiya:2016oao} in Sec.~\ref{sec:correction_central}.

\subsection{Validity and precision of weak-binding relation}
\label{sec:estimation}

We now propose the validity condition for the applicability of the weak-binding relation. For the validity condition to be satisfied, the exact value of the compositeness $X_{\rm exact}$ should be contained in the estimated uncertainty region of $X$. This condition can be expressed as
\begin{align}
\bar{X}_{l}(\xi)\leq X_{\rm exact}\leq \bar{X}_{u}(\xi).
\label{eq:validity}
\end{align}
For the previous weak-binding relation~\eqref{eq:wbr_s} to be applicable, it is necessary that the validity condition with $\xi=\xi_{\rm int}$ is satisfied. However, the improved relation is applicable when the validity condition with $\xi=\xi_{\rm int}$ ($\xi=\xi_{\rm eff}$) is satisfied for $\xi_{\rm int}>\xi_{\rm eff}$ ($\xi_{\rm int}<\xi_{\rm eff}$). Hence, we search for the applicable regions of the previous and improved weak-binding relations by examining the validity conditions with $\xi=\xi_{\rm int}$ and $\xi=\xi_{\rm eff}$. In particular, only the improved weak-binding relation is applicable when only the validity condition by $\xi_{\rm eff}$ holds for $\xi_{\rm eff}>\xi_{\rm int}$.

We also consider the precision of the estimated compositeness. When the uncertainty of the compositeness  is too large, we cannot determine the internal structure of the states from the weak-binding relation, even if the validity condition~\eqref{eq:validity} is satisfied. To estimate the precision, we introduce $\bar{E}$ as the magnitude of the uncertainty of Eq.~\eqref{eq:Xfinal}:
\begin{align}
\bar{E}=\bar{X}_{u}-\bar{X}_{l}.
\label{eq:Ebar}
\end{align}
Given that $0\leq X\leq 1$, we require $\bar{E}\leq0.5$ for the meaningful estimation in addition to the validity condition~\eqref{eq:validity}. While $\bar{E}$ can be negative with $X_{l}>1$ or $X_{u}<0$, the validity condition is never satisfied in this case because of the definition of $X$. We do not need to consider the precision in such a case because the weak-binding relation is not applicable.

\subsection{Effective range model}
\label{sec:er_model_stable}

In this section, we study the applicability of the weak-binding relations in the effective range model [$\rho_{0}\neq 0$ and $g_{0}=0$ in Eq.~\eqref{eq:H_effectiveresonance}] with a finite cutoff $\Lambda$. This model has the bare parameters $\lambda_{0}$ and $\rho_{0}$. The scattering length $a_{0}$, the effective range $r_{e}$, and the radius of the bound state $R$ are calculated from $\lambda_{0}$, $\rho_{0}$, and $\Lambda$ as shown in Sec.~\ref{sec:eft}. The exact value of the compositeness is known as $X_{\rm exact}=1$ (see Sec.~\ref{sec:models-X}). In this model, $\xi_{\rm eff}$ and $\xi_{\rm int}$ are
\begin{align}
\xi_{\rm eff}&=\frac{R_{\rm eff}}{R}=\frac{|r_e|}{R},
\label{eq:xi-eff}\\
\xi_{\rm int}&=\frac{R_{\rm int}}{R}=\frac{1}{\Lambda R},
\label{eq:xi-int}
\end{align}
where we note that $R_{\rm int}$ is estimated by the inverse of the cutoff $\Lambda$. By varying the bare parameters $\lambda_0$, $\rho_0$, and the cutoff $\Lambda$, we can arbitrarily choose the effective range $r_{e}$ and the radius of the bound state $R$. With these $r_{e}$ and $R$, the quantities $\xi_{\rm eff}$ and $\xi_{\rm int}$ are obtained from Eqs.~\eqref{eq:xi-eff} and \eqref{eq:xi-int}. Therefore, we can examine the validity conditions by $\xi_{\rm eff}$ and $\xi_{\rm int}$ separately.

\begin{figure}
\centering
\includegraphics[width=0.4\textwidth]{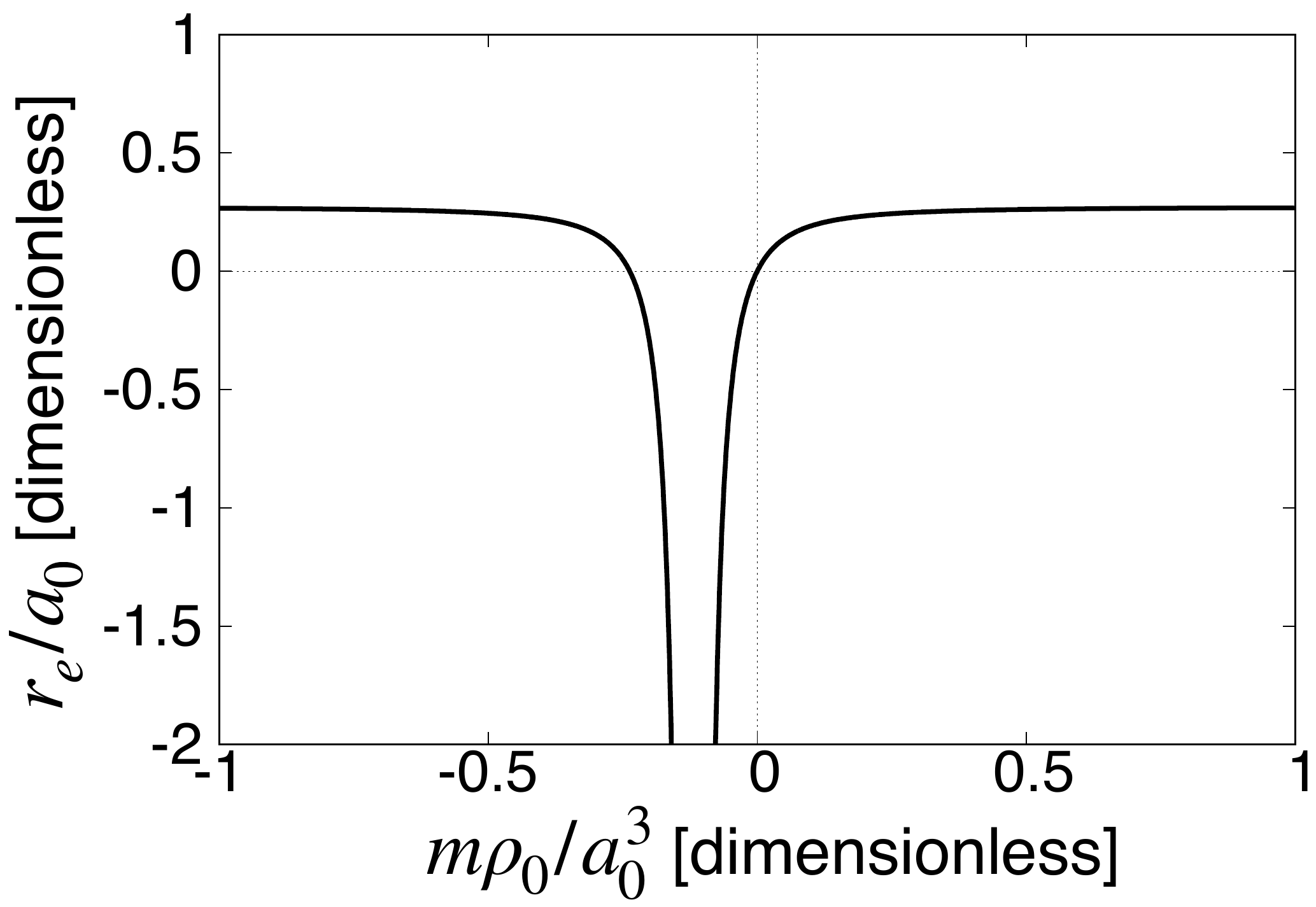}
\caption{The effective range $\tilde{r}_{e}=r_{e}/a_{0}$ [Eq.~\eqref{eq:tilde-re}] as a function of the bare parameter $\tilde{\rho}_{0}=m\rho_{0}/a_{0}^{3}$ with $\tilde{\Lambda}=a_{0}\Lambda=10$.}
\label{fig:re-rho0}
\end{figure}

For the convenience of numerical calculations, we introduce the dimensionless quantities. The dimensionless  $r_{e}$, $\rho_{0}$, and $\Lambda$ in units of $a_{0}$ and $m$ are given by
\begin{align}
\tilde{r}_{e}&=\frac{r_{e}}{a_{0}},\\
\tilde{\rho}_{0}&=\frac{m\rho_{0}}{a_{0}^{3}},\\
\tilde{\Lambda}&=a_{0}\Lambda.
\end{align}
We also define dimensionless length scales as $\tilde{R}_{\rm typ}=R_{\rm typ}/a_{0}$, $\tilde{R}_{\rm int}=R_{\rm int}/a_{0}$, and $\tilde{R}_{\rm eff}=R_{\rm eff}/a_{0}$. Because we choose $a_{0}$ as the unit of the length, the bare parameter $\lambda_{0}$ can be eliminated in Eq.~\eqref{eq:re-rho-lambda} with the dimensionless quantities. The effective range $\tilde{r}_{e}$ can be expressed by two variables $\tilde{\rho}_{0}$ and $\tilde{\Lambda}$ as
\begin{align}
\tilde{r}_{e}(\tilde{\rho}_{0},\tilde{\Lambda})&=\frac{2}{\pi^{3}}\frac{\tilde{\rho}_{0}[1+(1/24\pi^{2})\tilde{\rho}_{0}\tilde{\Lambda}^{3}]}{[1+(1/12\pi^{2})\tilde{\rho}_{0}\tilde{\Lambda}^{3}]^{2}}\left(\tilde{\Lambda}-\frac{\pi}{2}\right)^{2}.
\label{eq:tilde-re}
\end{align}
We plot $\tilde{r}_{e}$ as a function of $\tilde{\rho}_{0}$ with $\tilde{\Lambda}=10$ in Fig.~\ref{fig:re-rho0}. This plot shows that $|\tilde{r}_{e}|$ diverges at the pole of Eq.~\eqref{eq:tilde-re}:
\begin{align}
\tilde{\rho}_{0}=-\frac{12\pi^{2}}{\tilde{\Lambda}^{3}}.
\end{align}
We also find that $\tilde{r}_{e}$ is monotonically increasing for $\tilde{\rho}_{0}>-12\pi^{2}/{\tilde{\Lambda}^{3}}$:
\begin{align}
\frac{d\tilde{r}_{e}}{d\tilde{\rho}_0}&= \frac{2}{\pi^{3}}\frac{1}{[1+(1/12\pi^{2})\tilde{\rho}_{0}\tilde{\Lambda}^{3}]^{3}}\left(\tilde{\Lambda}-\frac{\pi}{2}\right)^{2} 
>0.
\label{eq:def-re}
\end{align}
Therefore, in the region $\tilde{\rho}_{0}>-12\pi^{2}/{\tilde{\Lambda}^{3}}$, there is one-to-one correspondence between  $\tilde{r}_{e}$ and $\tilde{\rho}_{0}$.
%
%
While the lower limit of $\tilde{r}_{e}$ is $-\infty$, the upper bound of $\tilde{r}_{e}$ [Eq.~\eqref{eq:tilde-re}] is given by the $\tilde{\rho}_{0}\to+\infty$ limit. Hence the values of $\tilde{r}_{e}$ are restricted in
\begin{align}
\label{eq:Wignerbound}
-\infty\leq \tilde{r}_{e}&\leq\frac{12}{\pi}\tilde{R}_{\rm int}-12\tilde{R}_{\rm int}^{2}+3\pi\tilde{R}_{\rm int}^{3}.
\end{align}
This upper bound is known as the Wigner bound~\cite{Matuschek:2020gqe}. We note that $\tilde{r}_{e}$ is always negative in the zero-range limit $\tilde{R}_{\rm int}\to 0$ because the Wigner bound becomes $\tilde{r}_{e}=0$.

\begin{figure*}[tbp]
\centering
\includegraphics[width=0.462\textwidth]{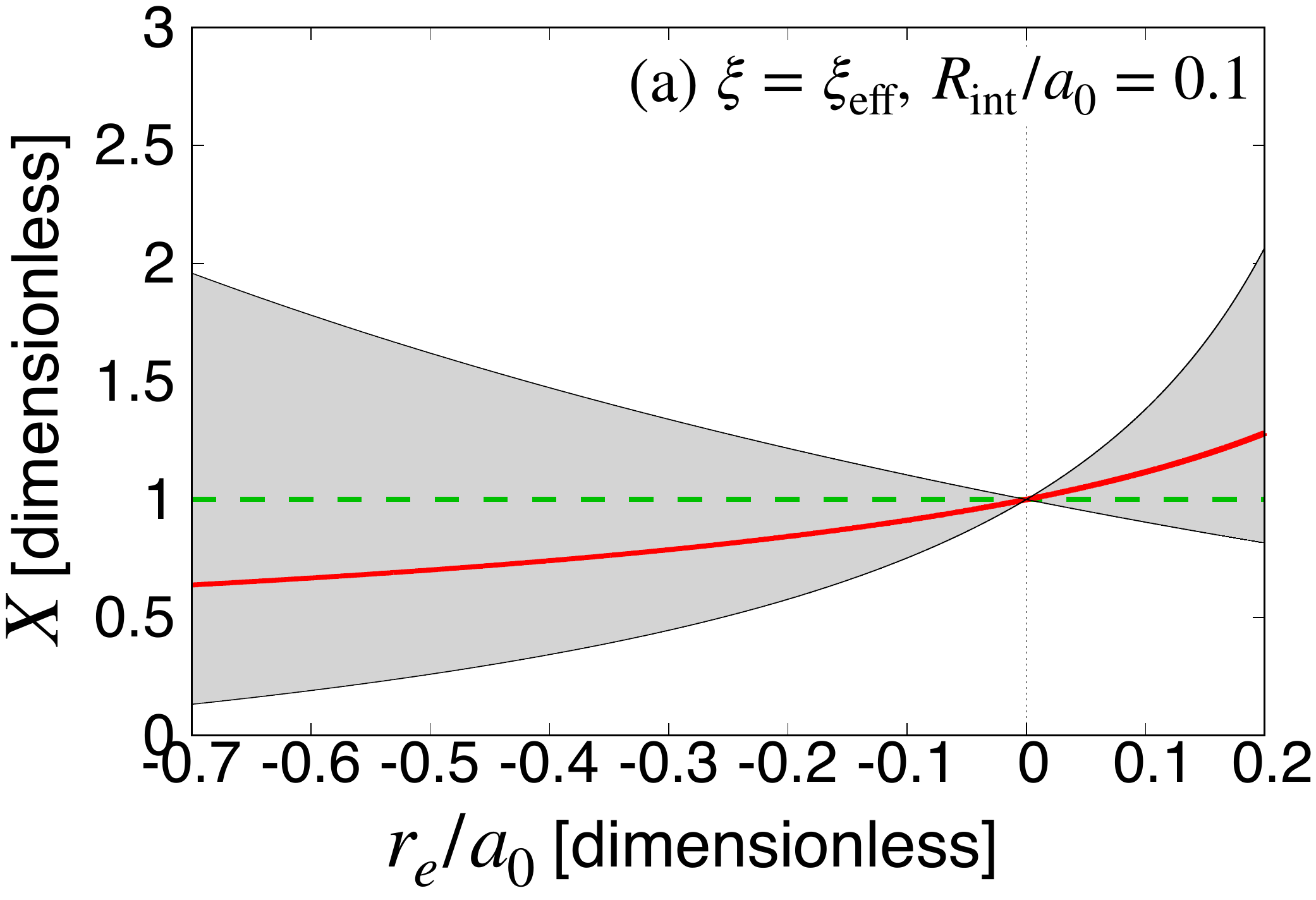}
\includegraphics[width=0.45\textwidth]{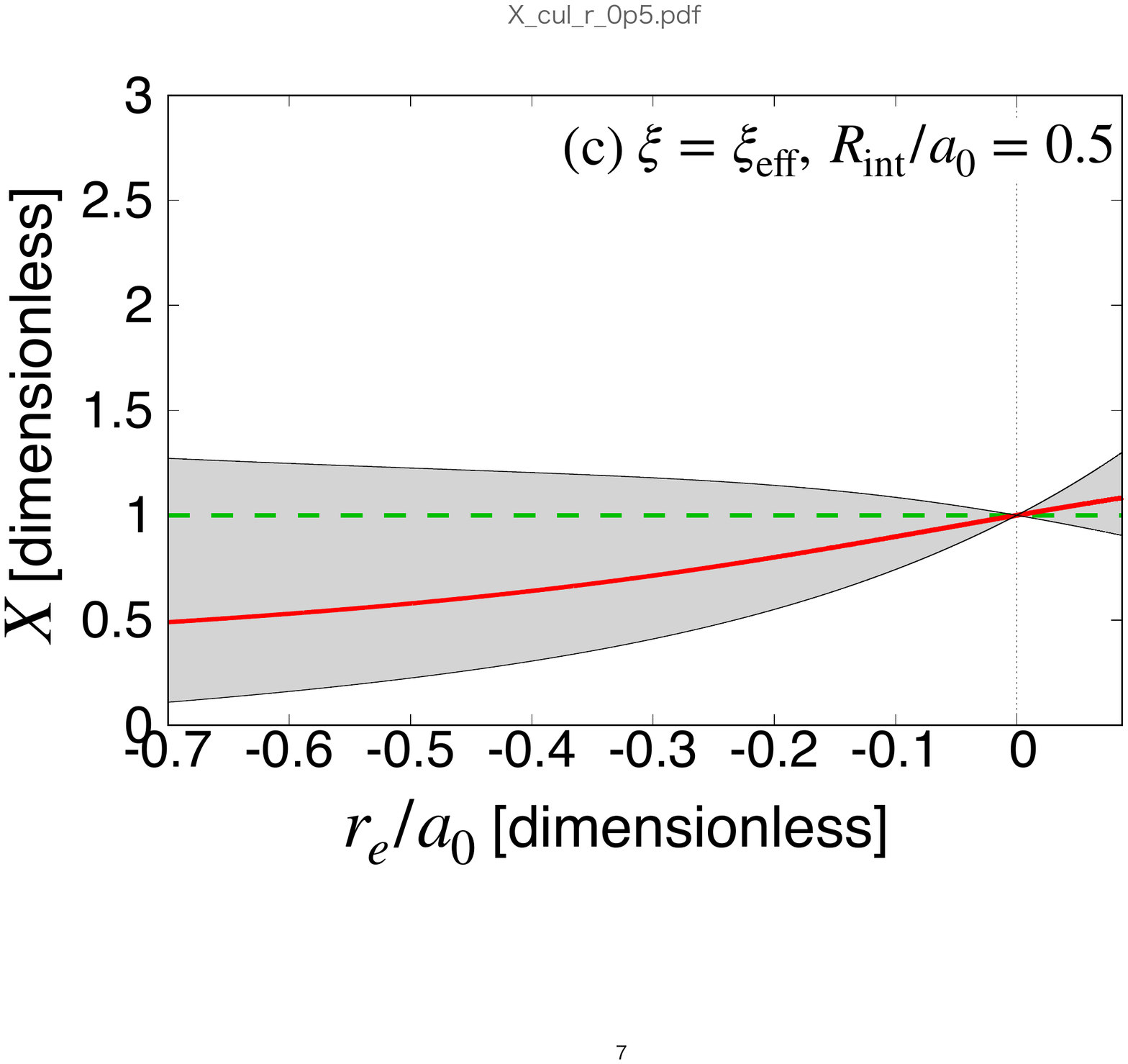}
\includegraphics[width=0.462\textwidth]{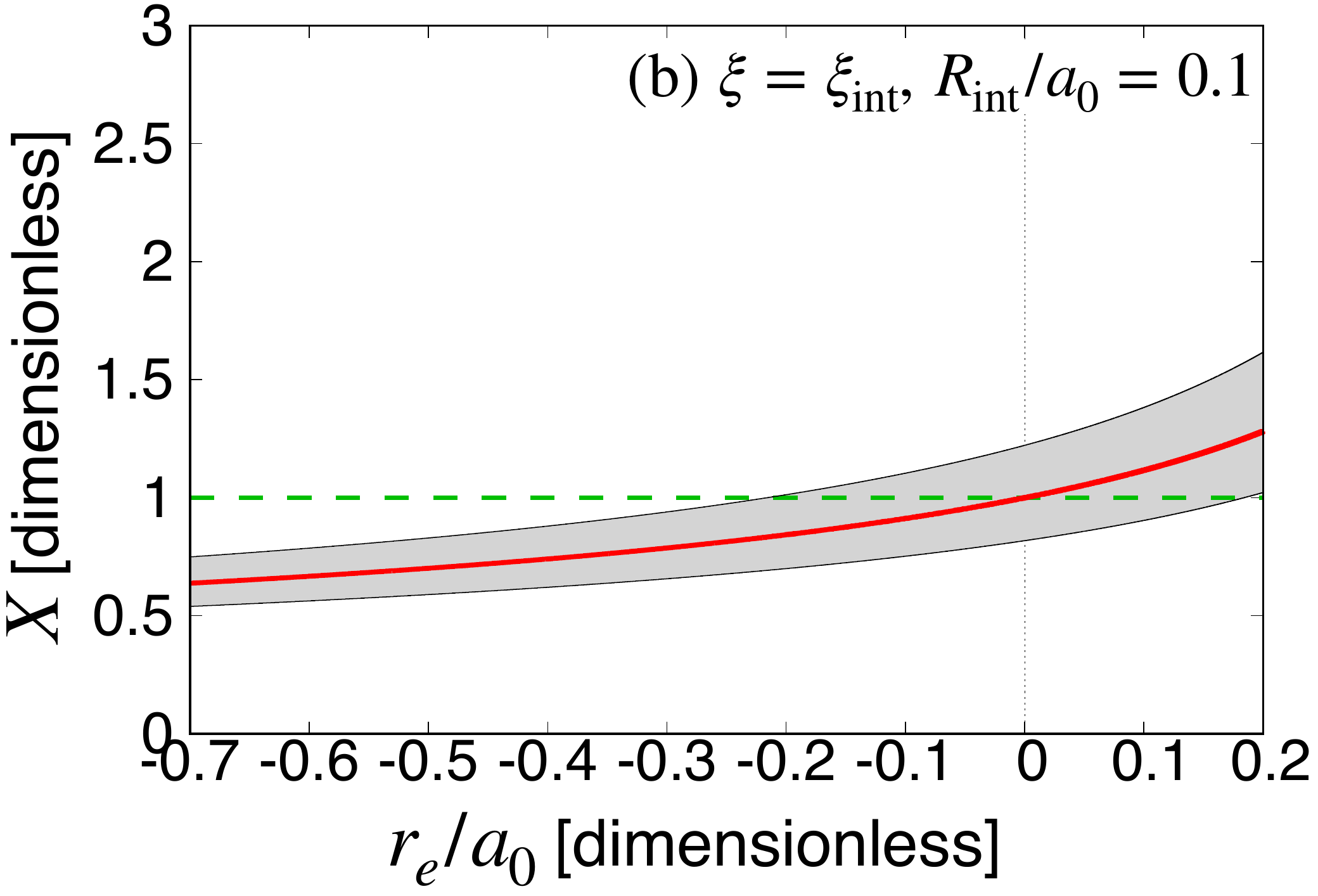}
\includegraphics[width=0.45\textwidth]{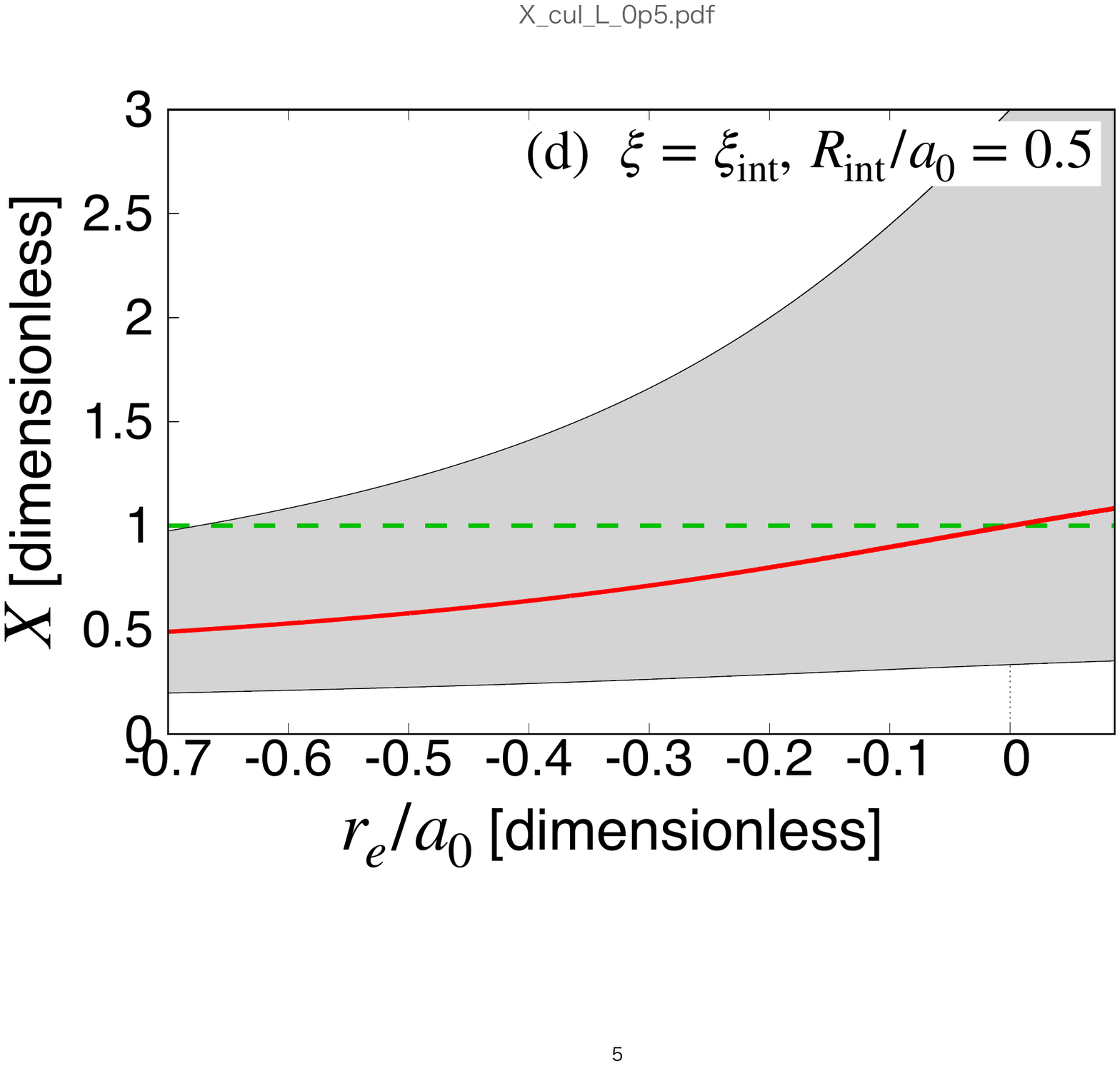}
\caption{The comparison of the exact value of the compositeness $X_{\rm exact}$ and that estimated by the weak-binding relations with [(a), (b)] $\tilde{R}_{\rm int}=0.1$ and [(c), (d)] $\tilde{R}_{\rm int}=0.5$. The dashed lines represent $X_{\rm exact}=1$. The central values of the compositeness are shown by the solid lines with the uncertainty band as functions of $\tilde{r}_{e}$. Panels (a) and (c) [(b) and (d)] show the case with $\xi=\xi_{\rm eff}$ ($\xi=\xi_{\rm int}$). The vertical dotted lines stand for $r_{e}/a_{0}=0$.}
\label{fig:X_boundary}
\end{figure*}

To study the applicable region of the weak-binding relations, we first discuss the behavior of the compositeness by varying the effective range $\tilde{r}_{e}$ with a fixed interaction range $\tilde{R}_{\rm int}$. We plot the central values of the compositeness $X_{c}$ (solid lines), the uncertainty bands in Eq.~\eqref{eq:uncertainty-band} (gray bands), and the exact values of the compositeness $X_{\rm exact}$ (dashed lines) as the functions of $\tilde{r}_{e}$ with $\tilde{R}_{\rm int}=0.1$ [Figs.~\ref{fig:X_boundary}(a) and \ref{fig:X_boundary}(b)] and with $\tilde{R}_{\rm int}=0.5$ [Figs.~\ref{fig:X_boundary}(c) and \ref{fig:X_boundary}(d)]. This figure is plotted slightly below the Wigner bound ($\tilde{r}_{e}\approx 0.27$ with $\tilde{R}_{\rm int}=0.1$ and $\tilde{r}_{e}\approx 0.09$ with $\tilde{R}_{\rm int}=0.5$). Figures \ref{fig:X_boundary}(a) and \ref{fig:X_boundary}(c) show the case with $\xi=\xi_{\rm eff}$, and Figs.~\ref{fig:X_boundary}(b) and \ref{fig:X_boundary}(d) with $\xi=\xi_{\rm int}$. In this figure, the central value $X_{c}$ [Eq.~\eqref{eq:Xc}] deviates from $X_{\rm exact}=1$ with the increase of $|\tilde{r}_{e}|$. In Figs.~\ref{fig:X_boundary}(a) and \ref{fig:X_boundary}(c), the uncertainty bands by $\xi_{\rm eff}$ also increase, and therefore the exact value $X_{\rm exact}=1$ is always contained in the uncertainty bands by $\xi_{\rm eff}$. In other words, the validity condition~\eqref{eq:validity} is always satisfied with $\xi=\xi_{\rm eff}$. We check that the same tendencies hold with other $\tilde{R}_{\rm int}\leq 0.5$, and the validity condition by $\xi_{\rm eff}$ is always satisfied in this region. However, in Figs.~\ref{fig:X_boundary}(b) and \ref{fig:X_boundary}(d), $X_u$ becomes smaller than $X_{\rm exact}$ at a large $|\tilde{r}_{e}|$. For example, $X_{\rm exact}$ is not contained in the region $\tilde{r}_{e}\lesssim -0.2$ with $\tilde{R}_{\rm int}=0.1$ [Fig.~\ref{fig:X_boundary}(b)]. Therefore, the validity condition by $\xi_{\rm int}$ is not satisfied in the large-$|\tilde{r}_{e}|$ region. This indicates that $X$ cannot be estimated by the previous weak-binding relation with $R_{\rm typ}=R_{\rm int}$ when the system has a large $|\tilde{r}_{e}|$. 

\begin{figure}[t]
\centering
\includegraphics[width=0.45\textwidth]{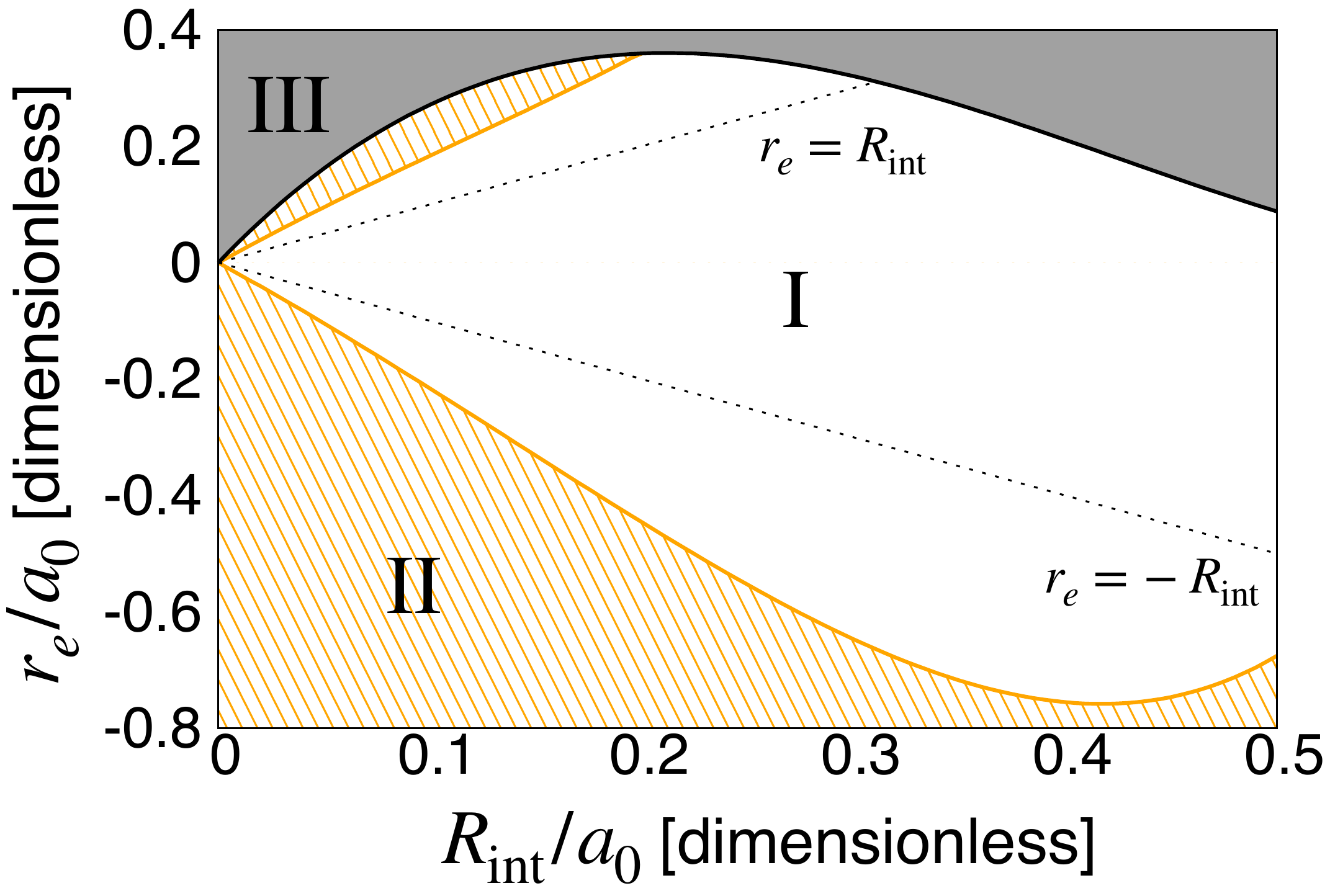}
\caption{The applicable regions of the weak-binding relations in the $\tilde{R}_{\rm int}$-$\tilde{r}_{e}$ plane. The solid lines which divide regions I and II are the boundaries at which the validity condition by $\xi_{\rm int}$ is violated. The lower bound of region III is the Wigner bound~\eqref{eq:Wignerbound}.  The dotted lines stand for $\tilde{r}_{e}=\pm \tilde{R}_{\rm int}$.}
\label{fig:parameter-regions}
\end{figure}

We now discuss the applicable regions of the weak-binding relations in the $\tilde{R}_{\rm int}$-$\tilde{r}_{e}$ plane. In Fig.~\ref{fig:parameter-regions}, we first plot the Wigner bound~\eqref{eq:Wignerbound} and denote the region above it as region III which is not realized in this model. The remaining region is divided into regions I and II by the boundary at which the validity condition by $\xi_{\rm int}$ is violated. The validity condition by $\xi_{\rm int}$ is satisfied in region I, but not in region II. The correction terms in the improved weak-binding relation~\eqref{eq:redef} are determined by the dotted lines $r_{e}=\pm R_{\rm int}$.  In region I, both the previous and improved relations are applicable, because the validity conditions by $\xi_{\rm eff}$ and $\xi_{\rm int}$ are satisfied. However, in region II, because the validity condition by $\xi_{\rm int}$ is violated, the previous weak-binding relation with $R_{\rm typ}=R_{\rm int}$ cannot be used. In region II, however, the improved relation~\eqref{eq:redef} is applicable because the validity condition by $\xi_{\rm eff}$ is satisfied and $R_{\rm typ}=R_{\rm eff}$ for $R_{\rm eff}>R_{\rm int}$. From these discussions, we find that the applicable region of the improved weak-binding relation (region ${\rm I}+{\rm II}$) is larger than the previous relation (region~I).

\begin{figure*}[tbp]
\centering
\includegraphics[width=0.45\textwidth]{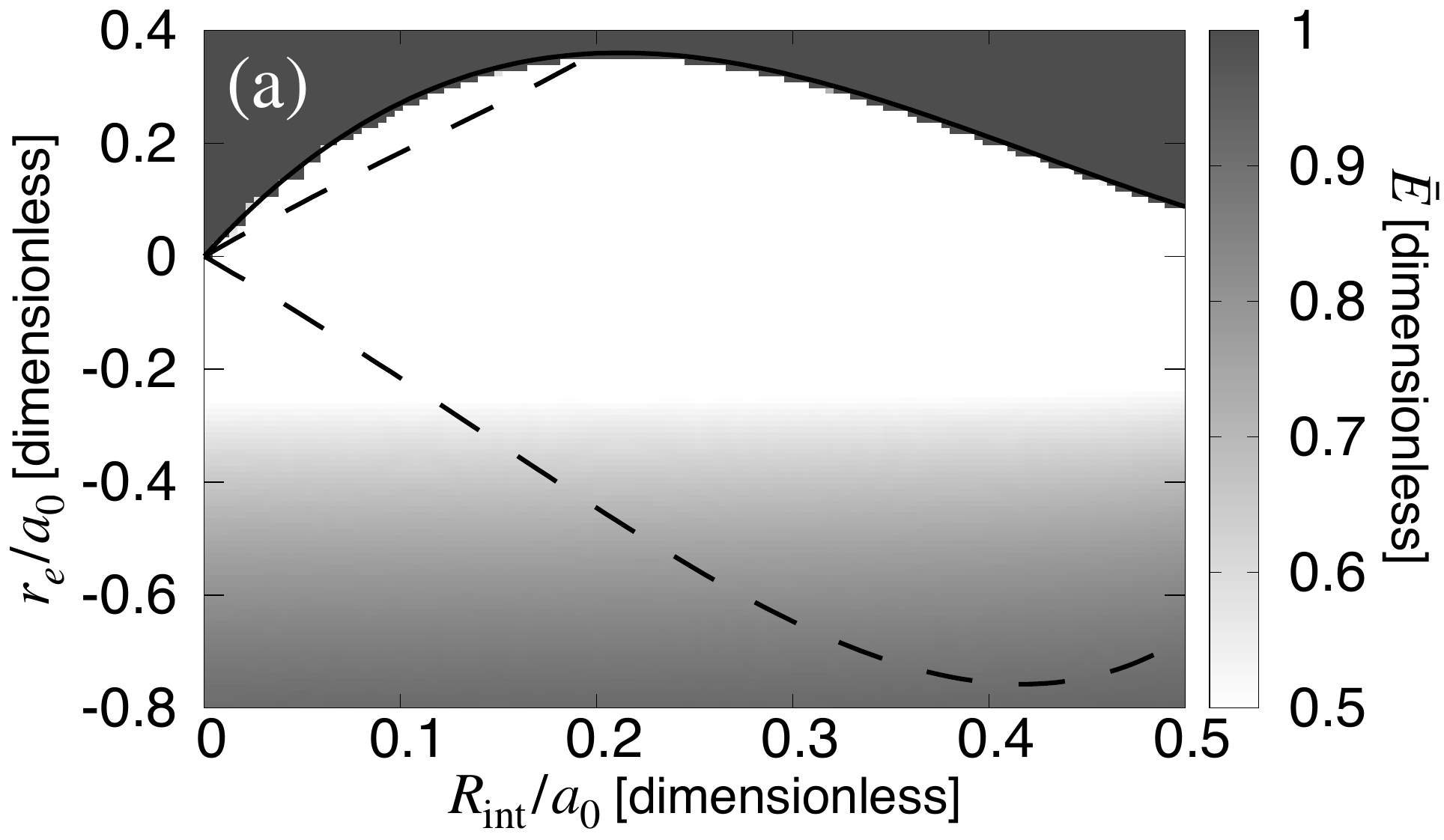}
\includegraphics[width=0.45\textwidth]{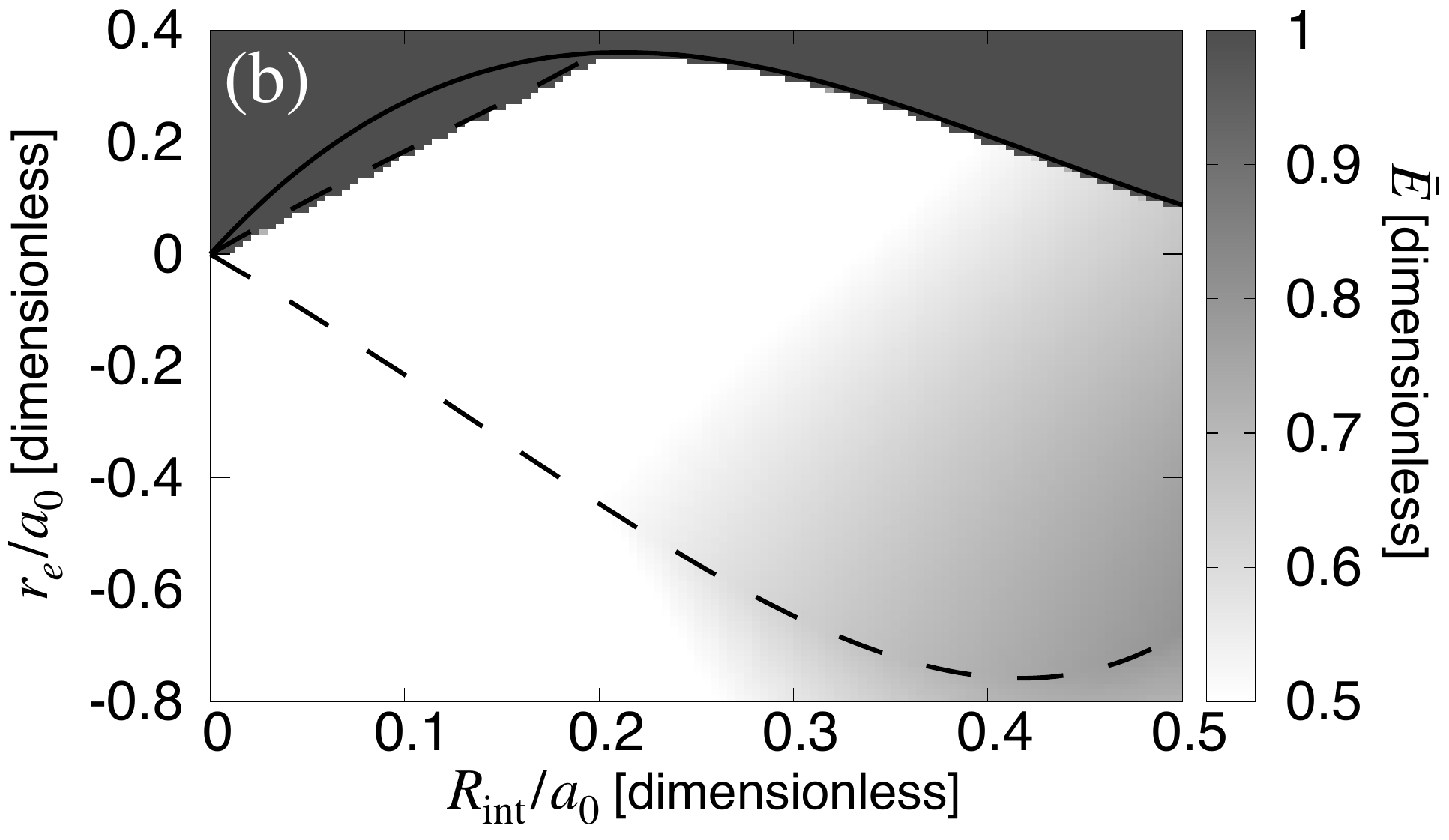}
\caption{The magnitude of the uncertainty $\bar{E}$ by (a) $\xi_{\rm eff}$ and (b) $\xi_{\rm int}$ in the $\tilde{R}_{\rm int}$-$\tilde{r}_{e}$ plane. The solid lines express the Wigner bound, and the dashed lines express the boundary of the applicable region of the previous weak-binding relation. }
\label{tab:Ebar}
\end{figure*}

We also consider the precision of the estimated compositeness by calculating the magnitude of the uncertainty $\bar{E}$ [Eq.~\eqref{eq:Ebar}] in this model. In Fig.~\ref{tab:Ebar}, we show the magnitude of the uncertainty $\bar{E}$ by $\xi_{\rm eff}$ [Fig.~\ref{tab:Ebar}(a)] and by $\xi_{\rm int}$ [Fig.~\ref{tab:Ebar}(b)] in the $\tilde{R}_{\rm int}$-$\tilde{r}_{e}$ plane. The irrelevant regions (above the Wigner bound or $X_{l}>1$) are filled in black. The unfilled regions correspond to $\bar{E}\leq0.5$ where the meaningful estimation is possible. In Fig.~\ref{tab:Ebar}(b), the change of the behavior of $\bar{E}$ at the boundary of the applicable region of the previous weak-binding relation can be explained by the difference of the definition of $\bar{E}$: $\bar{E}=1-X_{l}$ above the boundary and $\bar{E}=X_{u}-X_{l}$ below the boundary. In Fig.~\ref{tab:Ebar}(a), $\bar{E}$ by $\xi_{\rm eff}$ becomes smaller by decreasing the magnitude of the effective range $|r_{e}|$. This is because the uncertainty band increases with $|r_{e}|$ in Figs.~\ref{fig:X_boundary}(a) and \ref{fig:X_boundary}(c). Figure~\ref{tab:Ebar}(a) also shows that $\bar{E}$ is stable against the variation of $\tilde{R}_{\rm int}$. In fact, comparing with the uncertainty bands at the same $\tilde{r}_{e}$ in Figs.~\ref{fig:X_boundary}(a) and \ref{fig:X_boundary}(c), we find that $\bar{E}=1-X_{l}$ is not so sensitive to $R_{\rm int}$. 
When we compare $\bar{E}$ at positive and negative $\tilde{r}_{e}$ with a fixed $\tilde{R}_{\rm int}$, $\bar{E}$ with positive $\tilde{r}_{e}$ is smaller. For example, $\bar{E}\approx 0.4$ for $(\tilde{R}_{\rm int},\tilde{r}_{e})\approx(0.2,0.2)$ and $\bar{E}\approx 0.2$ for $(\tilde{R}_{\rm int},\tilde{r}_{e})\approx(0.2,-0.2)$. This reason is shown in Figs.~\ref{fig:X_boundary}(a) and \ref{fig:X_boundary}(c). In the $\tilde{r}_{e}>0$ region, a large part of the uncertainty band exceeds unity because the central value $X_{c}>1$, and therefore the remaining $\bar{E}$ is small. In contrast, in the $\tilde{r}_{e}<0$ region, the greater part of the uncertainty band is contained in the domain of $X$ because $X_{c}<1$. Hence, $\bar{E}$ is larger in the $\tilde{r}_{e}<0$ region.
In the Fig.~\ref{tab:Ebar}(b), $\bar{E}$ by $\xi_{\rm int}$ becomes smaller for the smaller $\tilde{R}_{\rm int}$ at the same $\tilde{r}_{e}$. In contrast to Fig.~\ref{tab:Ebar}(a), $\bar{E}$ by $\xi_{\rm int}$ depends not only on $\tilde{R}_{\rm int}$ but also on $\tilde{r}_{e}$. This reason is indicated by Figs.~\ref{fig:X_boundary}(b) and \ref{fig:X_boundary}(d); $\bar{E}$ varies with $\tilde{r}_{e}$ at the fixed $\tilde{R}_{\rm int}$.

\begin{figure}[tbp]
\centering
\includegraphics[width=0.45\textwidth]{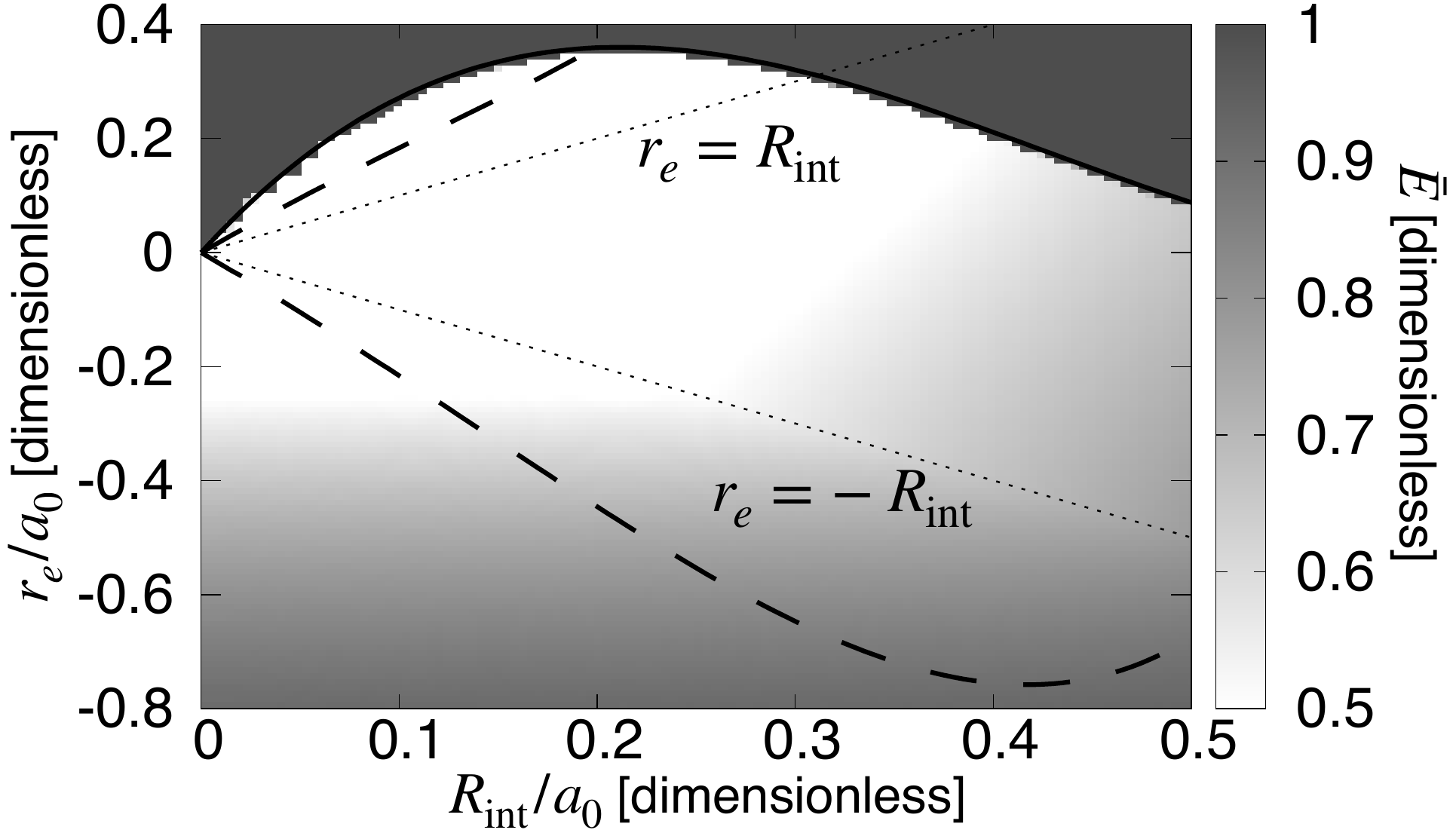}
\caption{The magnitude of the uncertainty for the weak-binding relation ($\bar{E}$ by $\xi_{\rm eff}$ in the $|\tilde{r}_{e}|>\tilde{R}_{\rm int}$ region, and by $\xi_{\rm int}$ in the $|\tilde{r}_{e}|<\tilde{R}_{\rm int}$ region) in the $\tilde{R}_{\rm int}$-$\tilde{r}_{e}$ plane. The solid lines express the Wigner bound, and the dashed lines express the boundary of the applicable region of the previous weak-binding relation. The dotted lines stand for $\tilde{r}_{e}=\pm \tilde{R}_{\rm int}$. }
\label{tab:Ebar-r-Lambda}
\end{figure}

In the improved weak-binding relation~\eqref{eq:redef}, the correction terms are determined by the values of the parameters $\tilde{r}_{e}$ and $\tilde{R}_{\rm int}$. Following the definition of $R_{\rm typ}$ [Eq.~\eqref{eq:redef}], we plot $\bar{E}$ by $\xi=\xi_{\rm eff}$ in $\tilde{r}_{e}>\tilde{R}_{\rm int}$ and $\tilde{r}_{e}<-\tilde{R}_{\rm int}$ and $\bar{E}$ by $\xi=\xi_{\rm int}$ in $-\tilde{R}_{\rm int}<\tilde{r}_{e}<\tilde{R}_{\rm int}$ in Fig.~\ref{tab:Ebar-r-Lambda}. We consider that the compositeness $X$ is estimated meaningfully for $\bar{E}\lesssim 0.5$ (unfilled region) as discussed in Sec.~\ref{sec:estimation}. In the $\tilde{R}_{\rm int}\lesssim 0.25$ region in Fig.~\ref{tab:Ebar-r-Lambda}, the meaningful estimation is possible in the $\tilde{r}_{e}\gtrsim -0.25$ region. In this region, the magnitude of the effective range $|\tilde{r}_{e}|$ is sufficiently smaller than the scattering length. The boundary for the meaningful estimation is determined only by $\tilde{r}_{e}$, because the uncertainty is estimated by $\xi_{\rm eff}$ there. In the $0.25\lesssim\tilde{R}_{\rm int}\lesssim 0.4$ region, the boundary for the meaningful estimation is contained in the $-\tilde{R}_{\rm int}<\tilde{r}_{e}<\tilde{R}_{\rm int}$ region and depends on both $\tilde{r}_{e}$ and $\tilde{R}_{\rm int}$. The reason for this behavior was discussed above for $\bar{E}$ estimated by $\xi_{\rm int}$. In the $\tilde{R}_{\rm int}\gtrsim0.4$ region, $\bar{E}$ is always larger than 0.5 and we cannot estimate $X$ meaningfully because the interaction range is not negligible in comparison with the scattering length. We conclude that the internal structure can safely be determined by the weak-binding relation in the $-0.25\lesssim\tilde{r}_{e}\lesssim0.25$ and $\tilde{R}_{\rm int}\lesssim 0.25$ region.

\subsection{Resonance model}
\label{sec:r-model-stable}

In the previous section, we consider the applicable region of the weak-binding relations in the effective range model ($\rho_{0}\neq 0$ and $g_{0}=0$) as a representative example with the exact value of the compositeness $X_{\rm exact}=1$. In this section, we also discuss the case with $X_{\rm exact}<1$, the resonance model ($\rho_{0}=0$ and $g_{0}\neq0$) with the finite cutoff $\Lambda$, introduced in Sec.~\ref{sec:eft}. As in Sec.~\ref{sec:er_model_stable} two kinds of uncertainties ($\xi_{\rm eff}$ and $\xi_{\rm int}$) are considered in this model because of the finite effective range~\eqref{eq:re-g-mu}.

The physical quantities are determined from the cutoff $\Lambda$ and three bare parameters $g_{0},\lambda_{0}$, and $\nu_{0}$ in this model. To reduce the number of the independent parameters, as in the effective range model, we introduce the dimensionless parameters in units of the scattering length $a_{0}$ and the boson mass $m$:
\begin{align}
\tilde{g}^{2}_{0}&=m^{2}a_{0}g_{0}^{2},
\label{eq:rm-tilde-g}\\
\tilde{\lambda}_{0}&=\frac{\lambda_{0}a_{0}}{m},
\label{eq:rm-tilde-lambda}\\
\tilde{\nu}_{0}&=ma_{0}^{2}\nu_{0},
\label{eq:rm-tilde-nu}\\
\tilde{R}&=\frac{R}{a_{0}},
\label{eq:tilde-R}\\
\tilde{\Lambda}&=a_{0}\Lambda,
\end{align}
where $\tilde{R}$ is determined by the pole condition of the scattering amplitude
\begin{align}
&\left[\left(1-\frac{2}{\pi}\tilde{\Lambda}\right)^{-1}+\frac{1}{8\pi}\frac{\tilde{g}_{0}^{2}}{\tilde{\nu}_{0}}+\frac{1}{8\pi}\frac{\tilde{g}_{0}^{2}}{{-\tilde{R}^{-2}}-\tilde{\nu}_{0}}\right]^{-1}\nonumber \\
&+\frac{2}{\pi}\tilde{\Lambda}-\frac{1}{\tilde{R}}=0.
\label{eq:rm-kappa}
\end{align}
We rewrite Eqs.~\eqref{eq:a0-g-mu} and \eqref{eq:re-g-mu} with the dimensionless parameters:
\begin{align}
1&=\left[8\pi\left(\tilde{\lambda}_{0}-\frac{\tilde{g}_{0}^{2}}{\tilde{\nu}_{0}}\right)^{-1}+\frac{2}{\pi}\tilde{\Lambda}\right]^{-1},
\label{eq:a0-g-mu-d}\\
\tilde{r}_{e}&=-\frac{16\pi \tilde{g}_{0}^{2}}{\tilde{\nu}_{0}^{2}}\left(\tilde{\lambda}_{0}-\frac{\tilde{g}_{0}^{2}}{\tilde{\nu}_{0}}\right)^{-2}.
\label{eq:re-g-mu-d}
\end{align}
By eliminating $\tilde{\lambda}_{0}$ with using Eq.~\eqref{eq:a0-g-mu-d}, the dimensionless effective range $\tilde{r}_{e}$ becomes:
\begin{align}
\tilde{r}_{e}&=-\frac{1}{4\pi}\frac{\tilde{g}_{0}^{2}}{\tilde{\nu}_{0}^{2}}\left(1-\frac{2}{\pi}\tilde{\Lambda}\right)^{2}.
\label{eq:rm-tilde-re}
\end{align}
We note that the Wigner bound is at $r_{e}=0$, because the effective range is always negative from Eq.~\eqref{eq:rm-tilde-re}. There are three independent parameters $\tilde{g}_{0}$, $\tilde{\nu}_{0}$, and $\tilde{\Lambda}$ in Eq.~\eqref{eq:rm-tilde-re}. To search for the applicable region in the $\tilde{r}_{e}$-$\tilde{R}_{\rm int}$ plane, we vary $\tilde{g}_{0}$ and $\tilde{\Lambda}$ for a fixed $\tilde{\nu}_{0}$ to determine $\tilde{r}_{e}$.

In contrast to the effective range model, the exact value of the compositeness $X_{\rm exact}$ depends on the parameters. The exact value of the compositeness is obtained as~\cite{Kamiya:2016oao}
\begin{align}
X_{\rm exact}
&=\frac{G'(-B,\Lambda)}{G'(-B,\Lambda)-[V^{-1}(-B)]'},
\label{eq:rm-X-exact}
\end{align}
where $B$ is the binding energy and $\alpha'(E)=d\alpha/d E$. In this model, the functions $V$ and $G$ are given by 
\begin{align}
V(E)&=\lambda_{0}+\frac{g_{0}^{2}}{E-\nu_{0}},
\label{eq:V}\\
G(E,\Lambda)&=\frac{m}{8\pi}\left(-\frac{2}{\pi}\Lambda-ik\right),
\label{eq:G}
\end{align}
where $E=k^{2}/m$ and the cutoff $\Lambda\gg 1/R$. By substituting Eqs.~\eqref{eq:V} and \eqref{eq:G} for  Eq.~\eqref{eq:rm-X-exact}, we obtain $X_{\rm exact}$
\begin{align}
X_{\rm exact}&=\left[1+\frac{16\pi\tilde{g}_{0}^{2}}{\tilde{R}\left\{\left(-\frac{1}{\tilde{R}^{2}}-\tilde{\nu}_{0}\right)\left(\frac{8\pi}{1-\frac{2}{\pi}\tilde{\Lambda}}+\frac{\tilde{g}_{0}^{2}}{\tilde{\nu}_{0}}\right)-\tilde{g}_{0}^{2}\right\}^{2}}\right]^{-1}.
\label{eq:rm-X-exact-2}
\end{align}
From this equation, we see that $X_{\rm exact}<1$ for $g_{0}\neq 0$, except for a special case with $\nu_{0}=0$ as discussed below.

We consider the applicable regions of the weak-binding relations in the $\tilde{r}_{e}$-$\tilde{R}_{\rm int}$ plane. We first discuss the special case with $\nu_{0}=\tilde{\nu}_{0}=0$ analytically. To consider the validity condition~\eqref{eq:validity}, we compare the central value $X_{c}$ with the exact value $X_{\rm exact}$. To this end, we take the limit $\tilde{\nu}_{0}\to 0$ in Eqs.~\eqref{eq:a0-g-mu-d} and \eqref{eq:re-g-mu-d} to obtain
\begin{align}
1&=\frac{\pi}{2\tilde{\Lambda}},\\
\tilde{r}_{e}&=-\frac{16\pi }{\tilde{g}_{0}^{2}}.
\end{align}
Here, we find that $\tilde{\Lambda}$ is constant with any $\tilde{g}_{0}$:
\begin{align}
\tilde{\Lambda}=\frac{\pi}{2}.
\label{eq:tilde-Lambda}
\end{align}
In other words, the interaction range is fixed in the $\tilde{\nu}_{0}=0$ case. $\tilde{R}$ is given from the pole condition~\eqref{eq:rm-kappa} with $\tilde{\nu}_{0}=0$: 
\begin{align}
\tilde{R}=\frac{\pi}{2\tilde{\Lambda}}.
\label{eq:tilde-kappa-rm}
\end{align}
Substituting Eq.~\eqref{eq:tilde-Lambda} for Eq.~\eqref{eq:tilde-kappa-rm}, we obtain $\tilde{R}=a_{0}/R=1$, and it is clear that $X_{c}=1$ from Eq.~\eqref{eq:Xc}. We also notice that $X_{\rm exact}=1$ from Eq.~\eqref{eq:rm-X-exact-2} with $\tilde{\nu}_{0}\to 0$. Therefore, we find that $X_{\rm exact}=X_{c}$, and the validity condition is always satisfied for any $\tilde{g}_{0}$ in the case with $\nu_{0}=0$. It is worth noting that we obtain $a_{0}=R$ even for the finite $\tilde{r}_{e}$ and $\tilde{\Lambda}$. Both $\tilde{r}_{e}$ and $\tilde{\Lambda}$ produce the deviation in the relation $a_{0}=R$ as discussed in Sec.~\ref{sec:rangecorrection}. To realize $a_{0}=R$, deviations by $\tilde{r}_{e}$ and $\tilde{\Lambda}$, including the higher order terms, should all cancel out. The resonance model in the limit $\tilde{\nu}_{0}\to0$ corresponds to such a special case.

\begin{figure*}
 \centering
\includegraphics[width=0.45\textwidth]{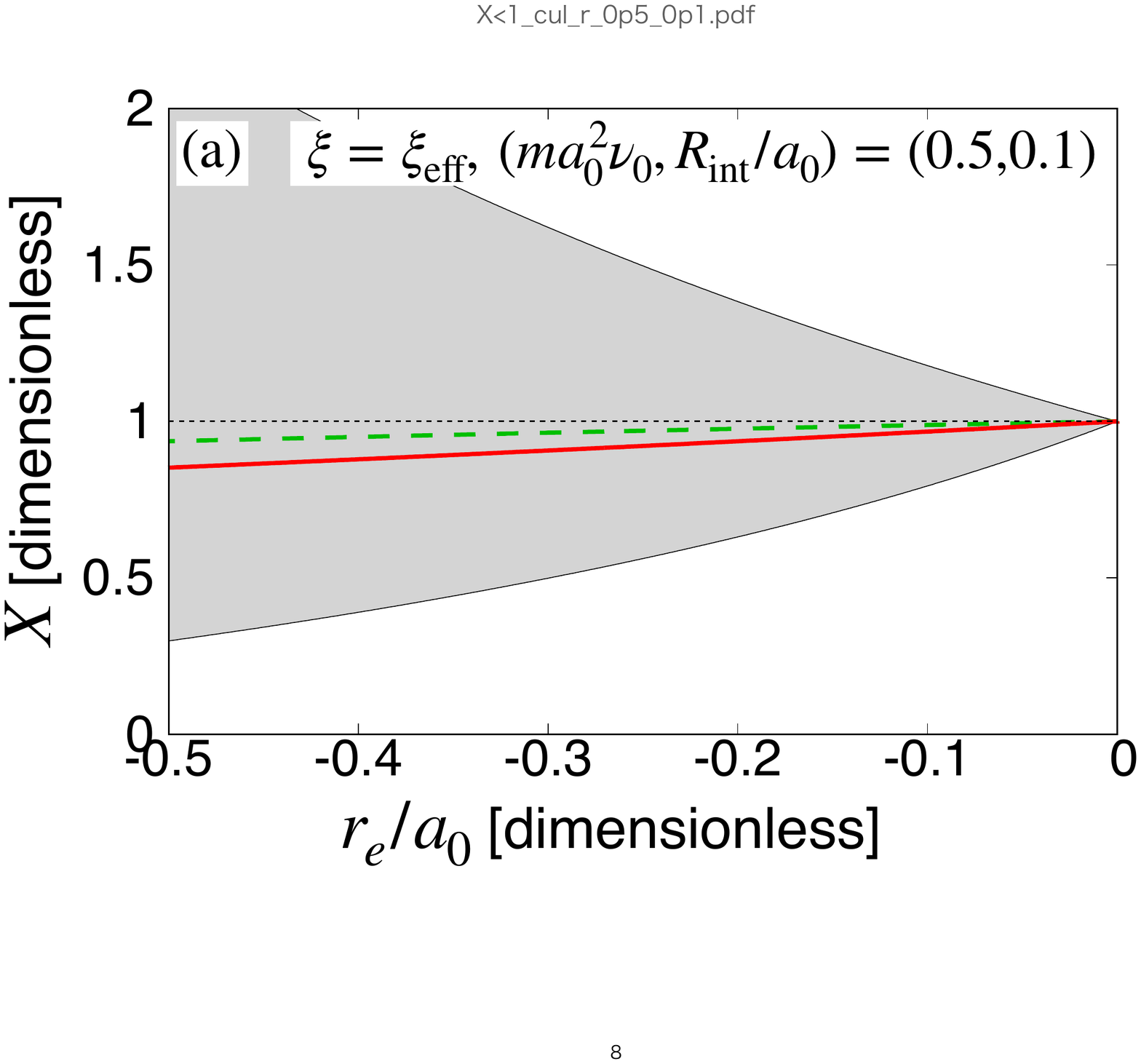}
\includegraphics[width=0.45\textwidth]{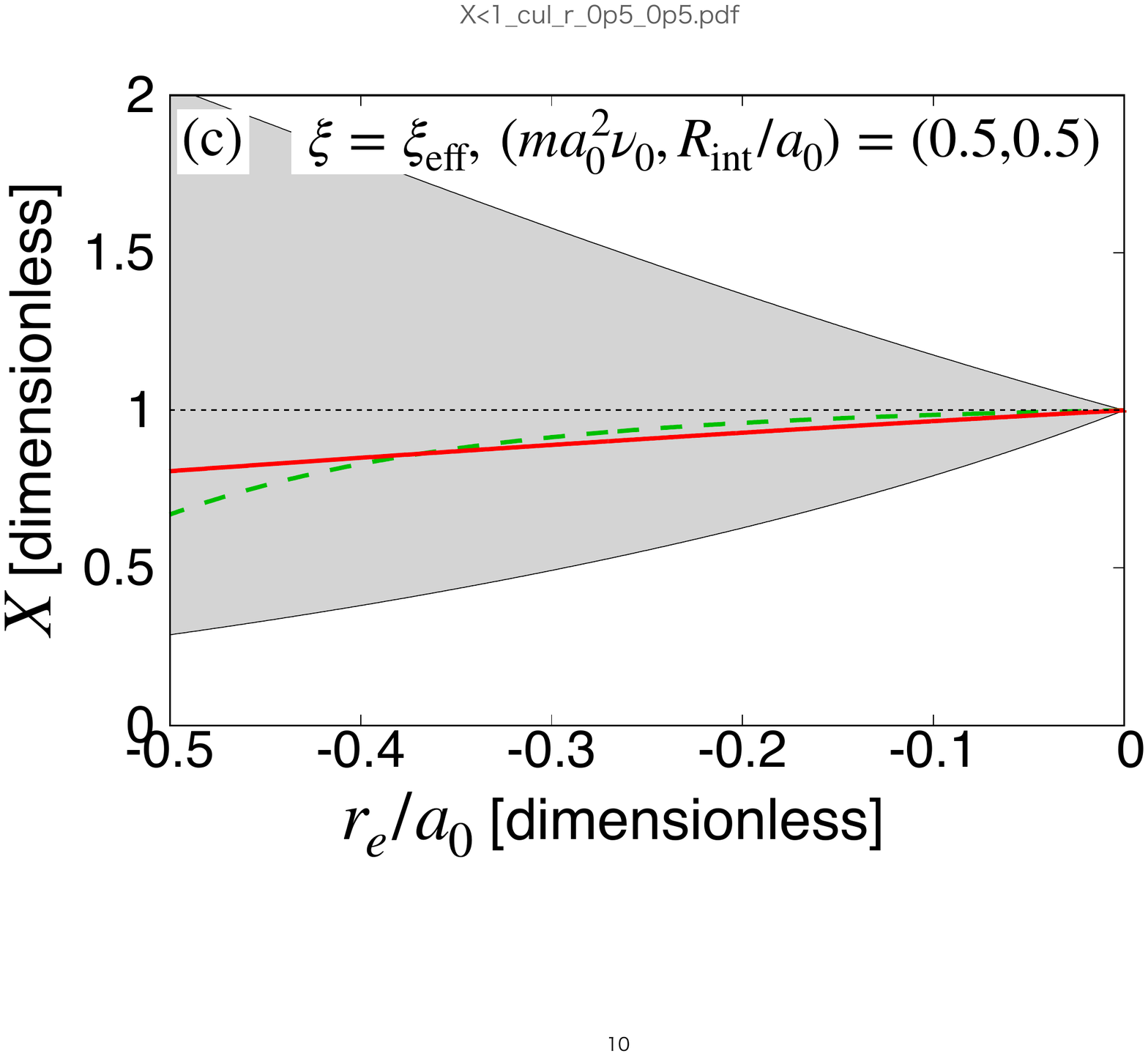}
\includegraphics[width=0.45\textwidth]{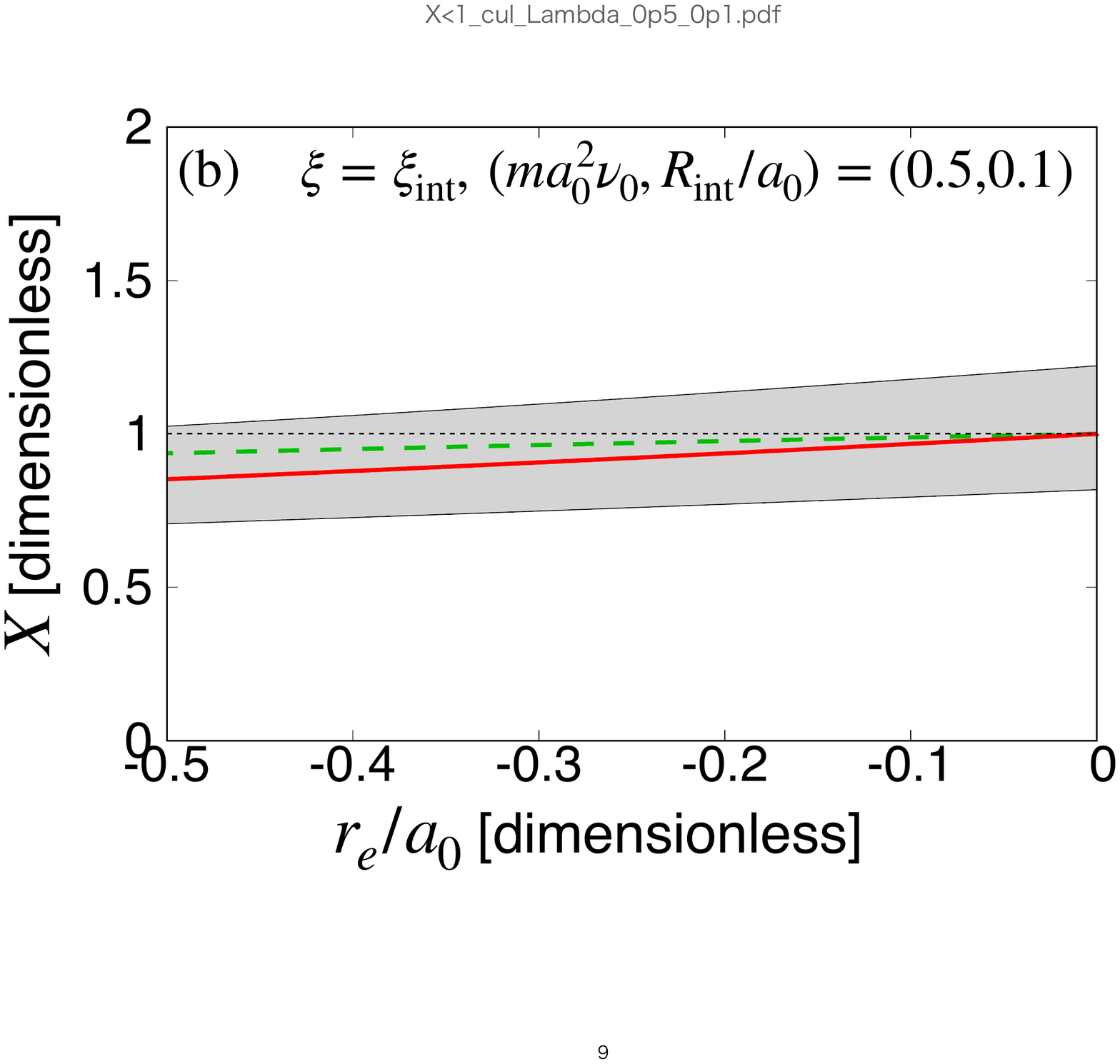}
\includegraphics[width=0.45\textwidth]{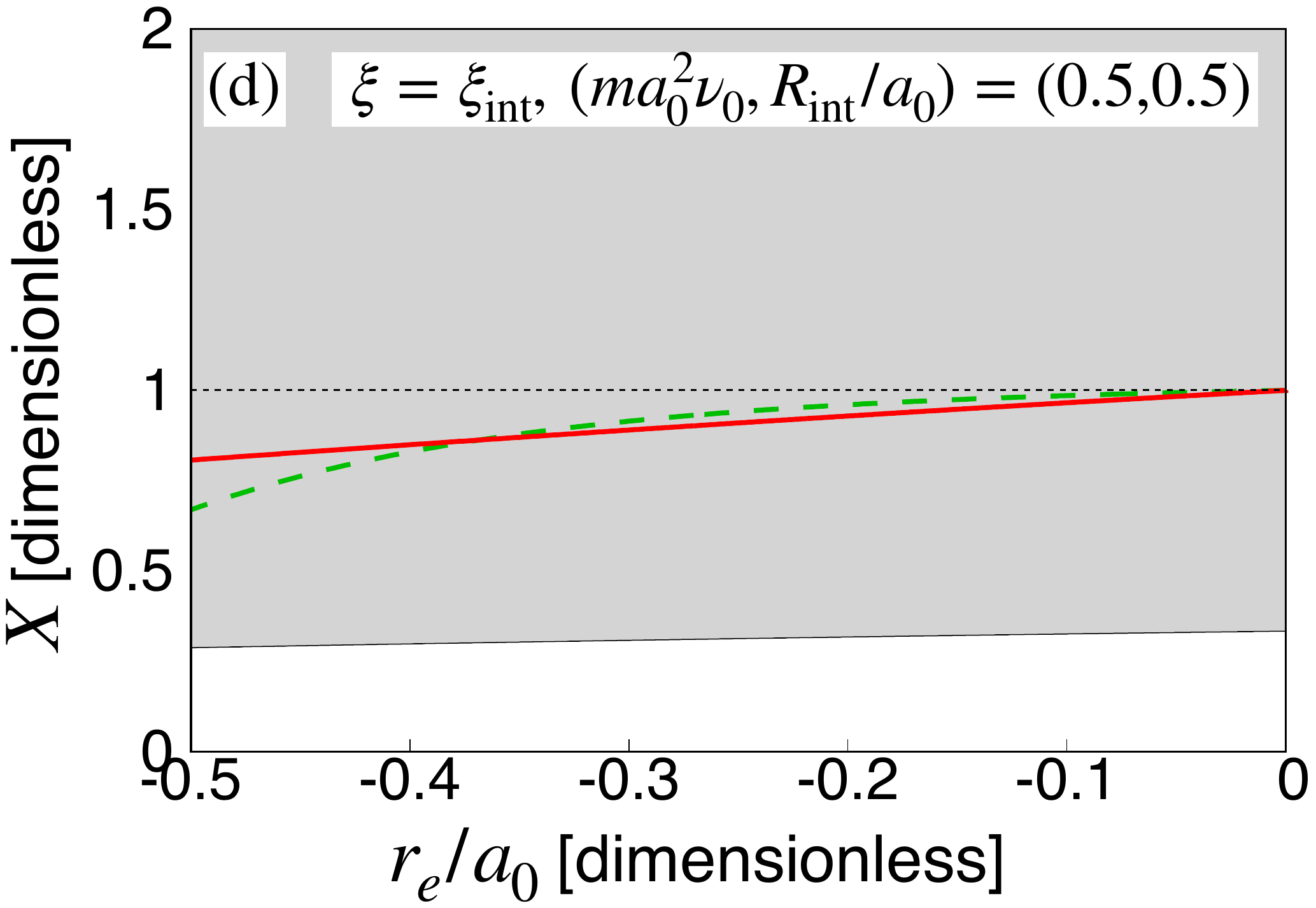}
 \caption{The comparison of the exact value of the compositeness $X_{\rm exact}$ and that estimated by the weak-binding relations with [(a) and (b)] $(\tilde{\nu}_{0},\tilde{R}_{\rm int})=(0.5,0.1)$ and [(c) and (d)] $(\tilde{\nu}_{0},\tilde{R}_{\rm int})=(0.5,0.5)$. The dashed lines represent $X_{\rm exact}$. The central values of the compositeness are shown by the solid lines with the uncertainty band as functions of $\tilde{r}_{e}$. The dotted lines stand for $X=1$. Panels (a) and (c) [(b) and (d)] show the case with $\xi=\xi_{\rm eff}$ ($\xi=\xi_{\rm int}$). }
 \label{fig:0p5}
 \end{figure*}

We then discuss the applicable regions of the weak-binding relations in the case with $\nu_{0}\neq0$ numerically by calculating the uncertainty band~\eqref{eq:uncertainty-band} and the exact value of the compositeness $X_{\rm exact}$~\eqref{eq:rm-X-exact-2}. To study the applicability in the $\tilde{R}_{\rm int}$-$\tilde{r}_{e}$ plane, we first examine the behavior of the uncertainty bands with $\xi_{\rm eff}$ and $\xi_{\rm int}$ as the functions of the effective range $\tilde{r}_{e}$ as in Sec.~\ref{sec:er_model_stable}. We plot the central values of the compositeness $X_{c}$ (solid lines), the uncertainty bands (gray bands) and the exact values of the compositeness $X_{\rm exact}$ (dashed lines) as the functions of $\tilde{r}_{e}$ with $\tilde{\nu}_{0}=0.5$, $\tilde{R}_{\rm int}=0.1$ [FIgs.~\ref{fig:0p5}(a) and \ref{fig:0p5}(b)] and with $\tilde{\nu}_{0}=0.5$, $\tilde{R}_{\rm int}=0.5$ [Figs.~\ref{fig:0p5} (c) and \ref{fig:0p5}(d)]. 
The dashed lines in those panels show that $X_{\rm exact}$ deviates from unity with the increase of $|\tilde{r}_{e}|$ in the resonance model. In Figs.~\ref{fig:0p5}(a) and \ref{fig:0p5}(c), because the uncertainty bands of $\xi_{\rm eff}$ become larger than the deviation of $X_{\rm exact}$ from unity, $X_{\rm exact}$ is always contained in the uncertainty region of $\xi_{\rm eff}$. However, the uncertainty band by $\xi_{\rm int}$ is basically determined by $\tilde{R}_{\rm int}$ and insensitive to $|\tilde{r}_{e}|$. It is therefore expected that there appears the region where $X_{\rm exact}$ is not contained in the uncertainty band for a sufficiently small $\tilde{R}_{\rm int}$. In fact, we find that $X_{\rm exact}$ goes beyond the uncertainty band in the $\tilde{r}_{e}\leq-0.5$ region with $\tilde{R}_{\rm int}\lesssim0.5$. This is the same tendency as the effective range model.

\begin{figure*}
 \centering
\includegraphics[width=0.45\textwidth]{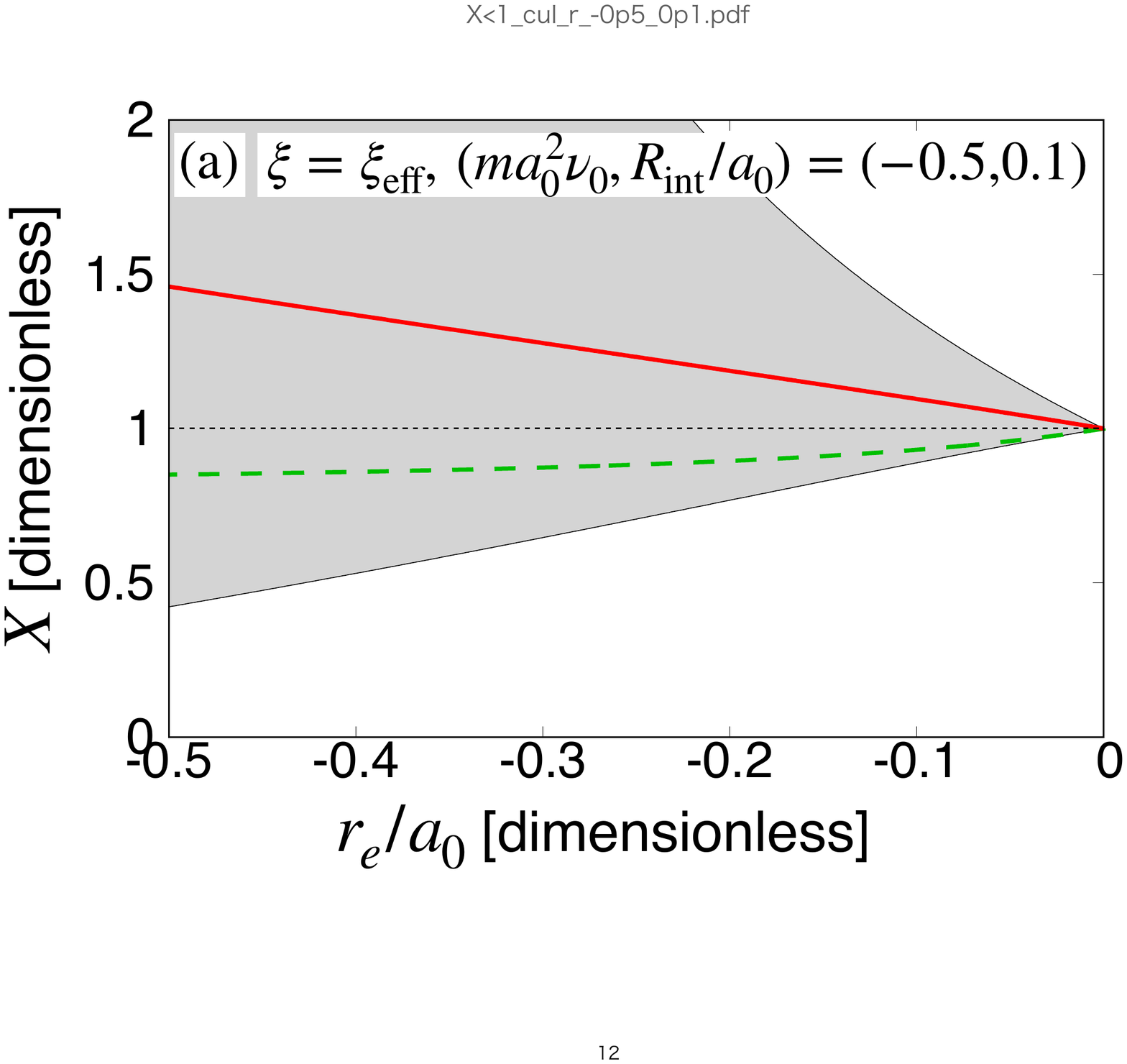}
\includegraphics[width=0.45\textwidth]{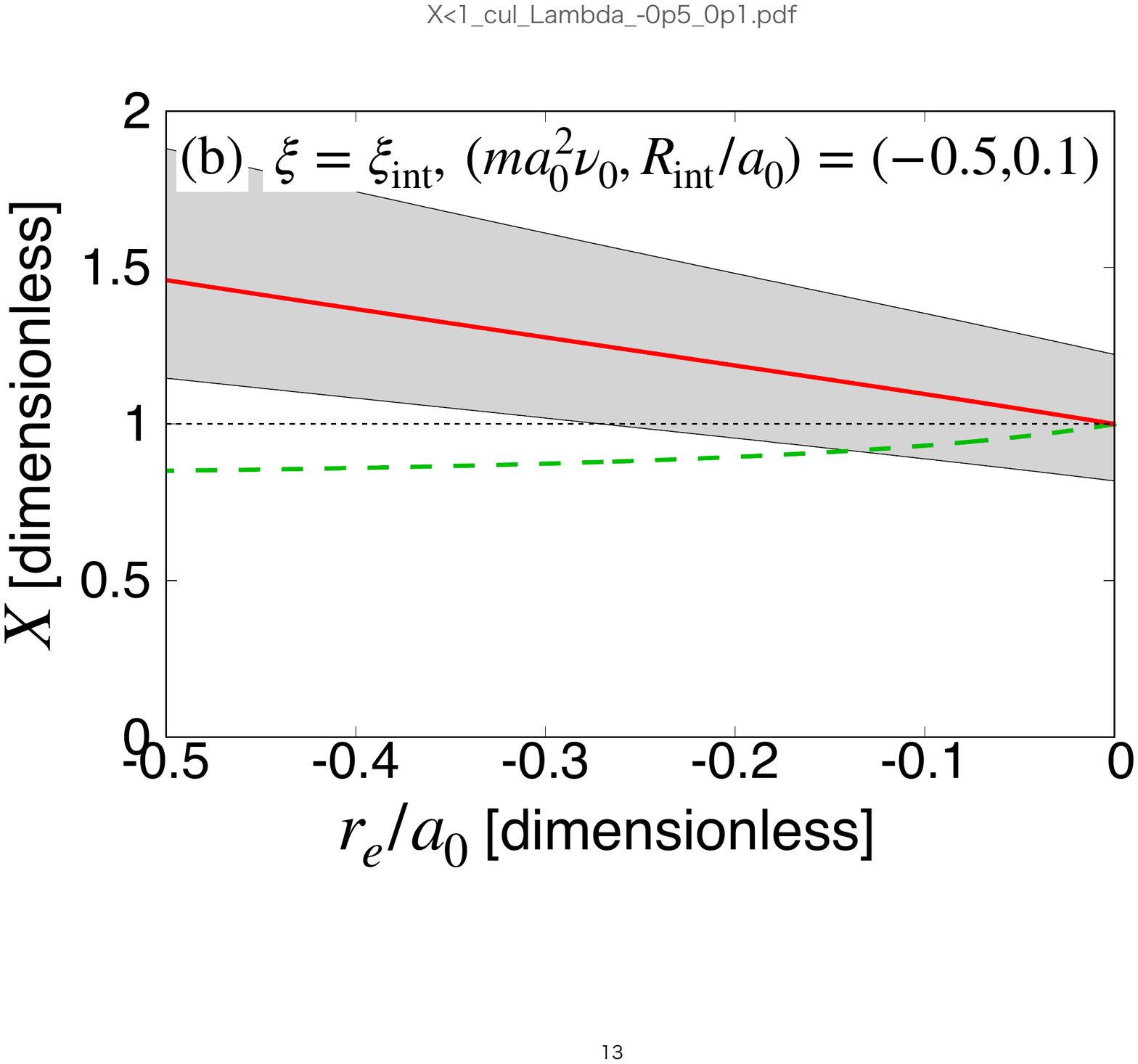}
 \caption{The comparison of the exact value of the compositeness $X_{\rm exact}$ and that estimated by the weak-binding relations with $(\tilde{\nu}_{0},\tilde{R}_{\rm int})=(-0.5,0.1)$. The dashed lines represent $X_{\rm exact}$. The central values of the compositeness are shown by the solid lines with the uncertainty band as functions of $\tilde{r}_{e}$. The dotted lines stand for $X=1$. (a) $\xi=\xi_{\rm eff}$, (b) $\xi=\xi_{\rm int}$. \label{fig:-0p5-0p1}}
 \end{figure*}

In Fig.~\ref{fig:-0p5-0p1}, we also plot the results with negative $\tilde{\nu}_{0}=-0.5$ and $\tilde{R}_{\rm int}=0.1$. While $X_{\rm exact}$ is smaller than unity, the central value $X_{c}$ is larger than unity in Fig.~\ref{fig:-0p5-0p1}. By comparing Fig.~\ref{fig:0p5}(b) with Fig.~\ref{fig:-0p5-0p1}, we find that the behaviors of $X_{c}$ are different from each other in contrast to $X_{\rm exact}$. Hence, it is expected that $X_{\rm exact}$ goes beyond the uncertainty band at a smaller $|\tilde{r}_{e}|$ with $\tilde{\nu}_{0}=-0.5$ than that with $\tilde{\nu}_{0}=0.5$. Indeed, the validity condition by $\xi_{\rm int}$ is not satisfied for $|\tilde{r}_{e}|\lesssim 0.13$ in Fig.~\ref{fig:-0p5-0p1}. However, we check that the validity condition by $\xi_{\rm eff}$ is always satisfied in the $-0.5\leq\tilde{r}_{e}\leq0$ region for $\tilde{R}_{\rm int}\leq0.5$ because the uncertainty band is sufficiently large.

\begin{figure*}
 \centering
\includegraphics[width=0.45\textwidth]{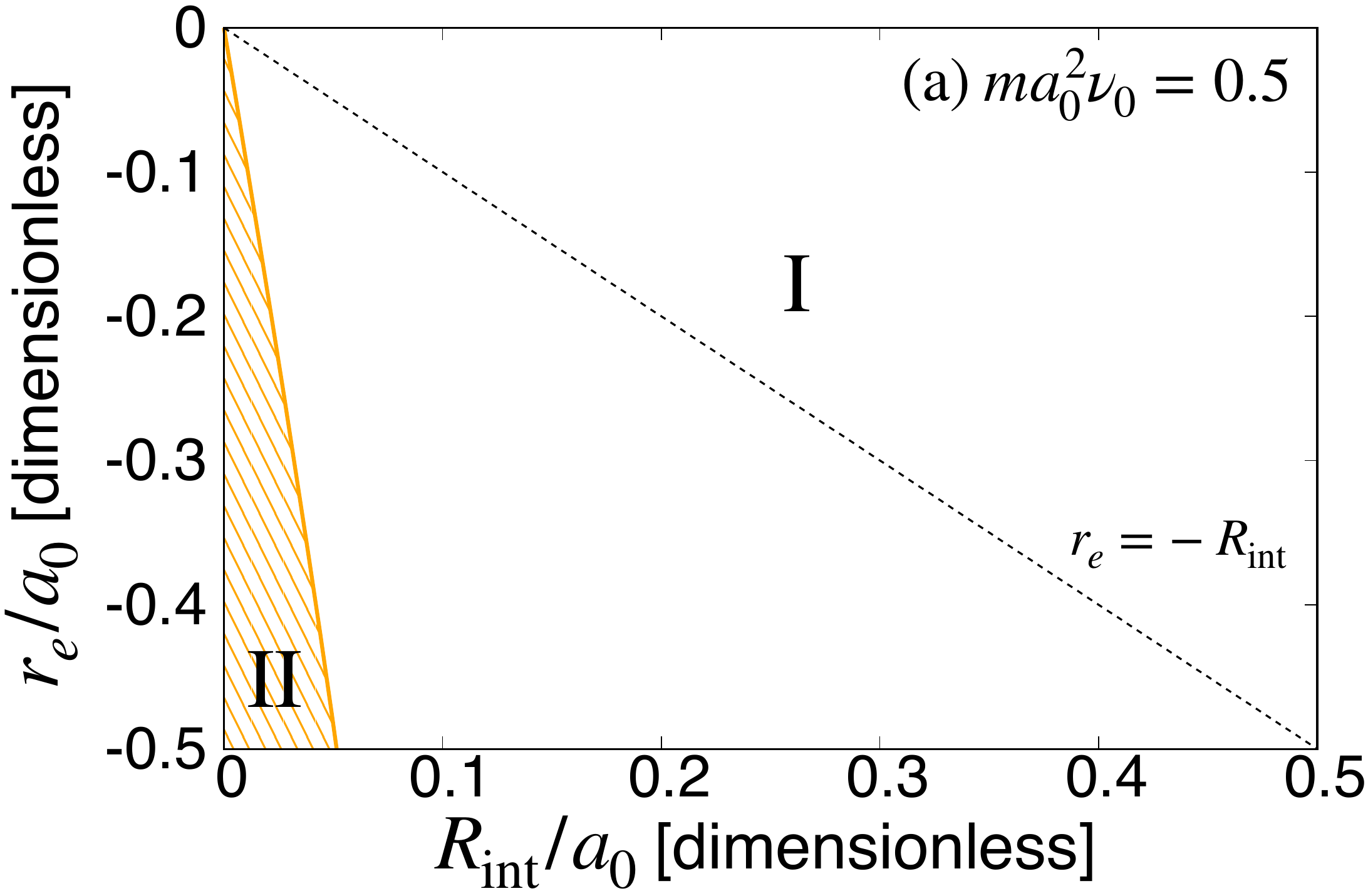}
\includegraphics[width=0.45\textwidth]{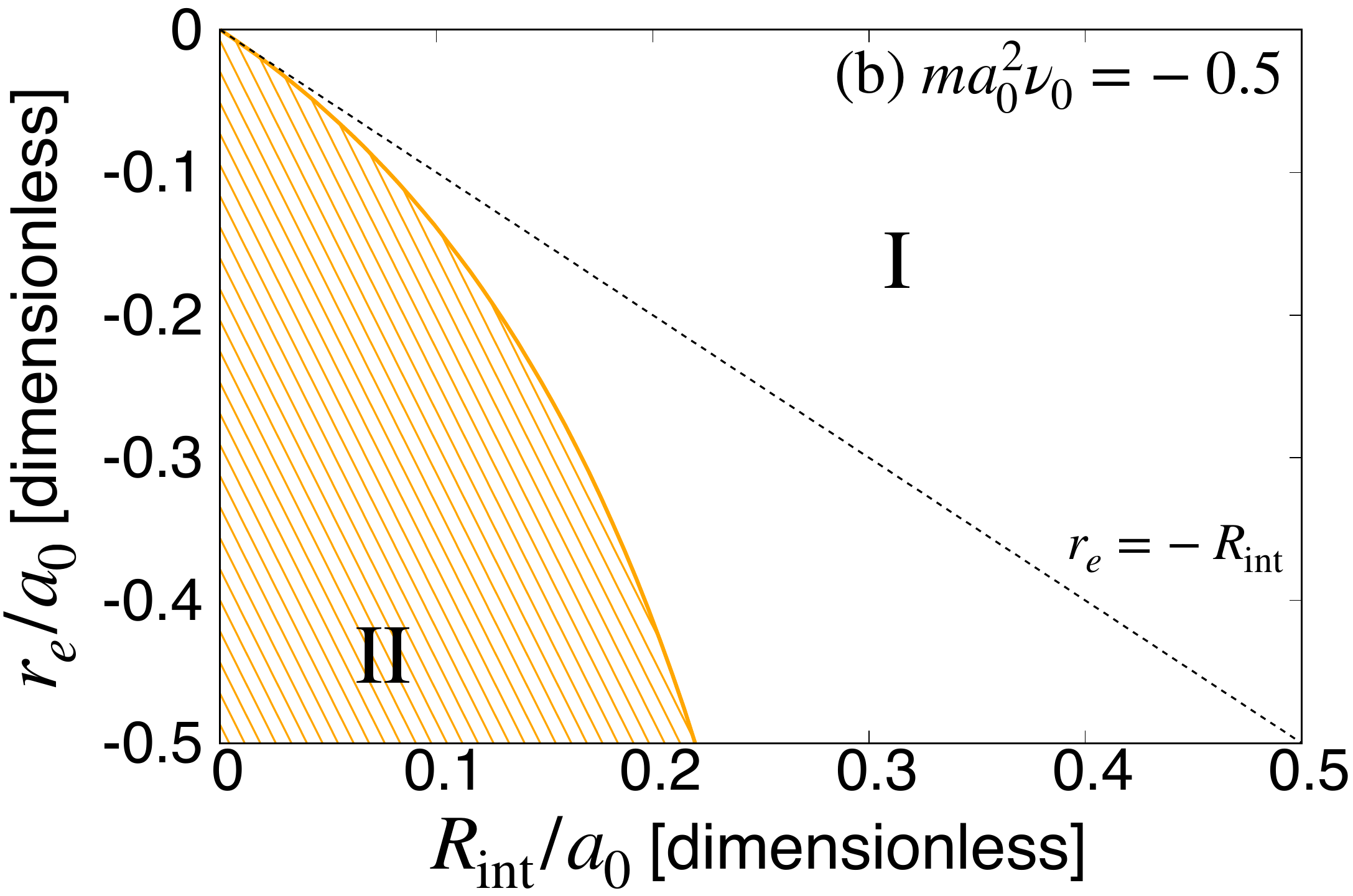}
\caption{The applicable regions of the weak-binding relations in the case with (a) $\tilde{\nu}_{0}=0.5$ (b) $\tilde{\nu}_{0}=-0.5$ in the $\tilde{R}_{\rm int}$-$\tilde{r}_{e}$ plane. The solid lines are the boundaries at which the validity condition by $\xi_{\rm int}$ is violated. }
\label{fig:regions_rm}
\end{figure*}

We now plot the applicable regions of the weak-binding relations in the $\tilde{R}_{\rm int}$-$\tilde{r}_{e}$ plane for the resonance model with $\tilde{\nu}_{0}=0.5$ [Fig.~\ref{fig:regions_rm}(a)] and $\tilde{\nu}_{0}=-0.5$ [Fig.~\ref{fig:regions_rm}(b)]. Those figures are plotted in the same manner as Fig.~\ref{fig:parameter-regions} for the effective range model. Both the previous and improved relations are applicable in region I where the validity conditions by $\xi_{\rm int}$ and $\xi_{\rm eff}$ are satisfied. In contrast, only the improved relation is applicable in region II where the validity condition by $\xi_{\rm int}$ is not satisfied. Hence, the applicable region becomes large (I$\to$I$+$II) with the range correction in the resonance model with $X_{\rm exact}<1$ as in the same way with the effective range model ($X_{\rm exact}=1$). By comparing the Figs.~\ref{fig:regions_rm}(a) and \ref{fig:regions_rm}(b), we find that region I for $\tilde{\nu}_{0}=-0.5$ is smaller than that for $\tilde{\nu}_{0}=0.5$. This can be understood by the above discussion on the behavior of the uncertainty bands and $X_{\rm exact}$ in Figs.~\ref{fig:0p5} and \ref{fig:-0p5-0p1}.

\subsection{Higher order terms of effective range expansion}
\label{sec:correction_central}

In this section, we numerically examine the improvement of the central value of the compositeness $X_{c}$ with the effective range expansion in the previous work~\cite{Kamiya:2016oao}. By eliminating the scattering length $a_{0}$ from Eq.~\eqref{eq:Xc} with the pole condition up to the second order of the momentum $k$ in the effective range expansion,
\begin{align}
-\frac{r_e}{2R^2}+\frac{1}{R}&=\frac{1}{a_0},
\label{eq:ere-2}
\end{align}
we obtain the central value of the compositeness expressed by the effective range $r_{e}$ and the radius $R$:
\begin{align}
X_{c}(r_e,R)&=\frac{R}{R-r_e}.
\label{eq:X-re-R-4}
\end{align}
To introduce the contributions of the higher order terms in the effective range expansion, we use $X$ expressed by $r_{e}$ and the shape parameter $P_{s}$ calculated from the residue of the bound-state pole in the scattering amplitude up to the fourth order of $k$~\cite{Kamiya:2016oao}:
\begin{align}
X_{c}(r_{e},P_{s})&=\left[1-\frac{r_{e}}{R}-\frac{P_{s}}{R}\right]^{-1}.
\label{eq:X-re-Ps}
\end{align}
By considering the pole condition up to $k^{4}$ in the effective range expansion
\begin{align}
-\frac{r_e}{2R^2}+\frac{1}{R}-\frac{1}{4}\frac{P_{s}}{R^4}&=\frac{1}{a_0},
\label{eq:ere-4}
\end{align}
$P_{s}$ can be eliminated from Eq.~\eqref{eq:X-re-Ps} to obtain 
\begin{align}
X_{c}(a_0,r_e,R)&=\left[\frac{4R}{a_0}+\frac{r_e}{R}-3\right]^{-1}.
\label{eq:X-a0-re-R-4}
\end{align}
In this expression, the effect of $P_{s}$ is implicitly included. We expand Eq.~\eqref{eq:X-re-Ps} in powers of $P_{s}/R^{3}$:
\begin{align}
X_{c}&=\frac{R}{R-r_{e}}\left(1+\frac{P_{s}}{R^3}\right)+\cdots.
\label{eq:X-Ps}
\end{align}
If $|P_{s}|/R^{3}\ll 1$, we can neglect the higher order terms of $P_{s}$, and Eq.~\eqref{eq:X-Ps} reduces to Eq.~\eqref{eq:X-re-R-4}. Because Eq.~\eqref{eq:X-re-R-4} is equivalent to Eq.~\eqref{eq:Xc} under the condition of Eq.~\eqref{eq:ere-2}, Eq.~\eqref{eq:X-a0-re-R-4} would be an improvement of the weak-binding relation with the effects of $P_{s}$. We note that the difference between Eqs.~\eqref{eq:Xc} and \eqref{eq:X-re-R-4} also expresses the effects of $P_{s}$ because its contribution is neglected in Eq.~\eqref{eq:ere-2}.

\begin{figure}[tbp]
  \centering
  \includegraphics[width=0.45\textwidth]{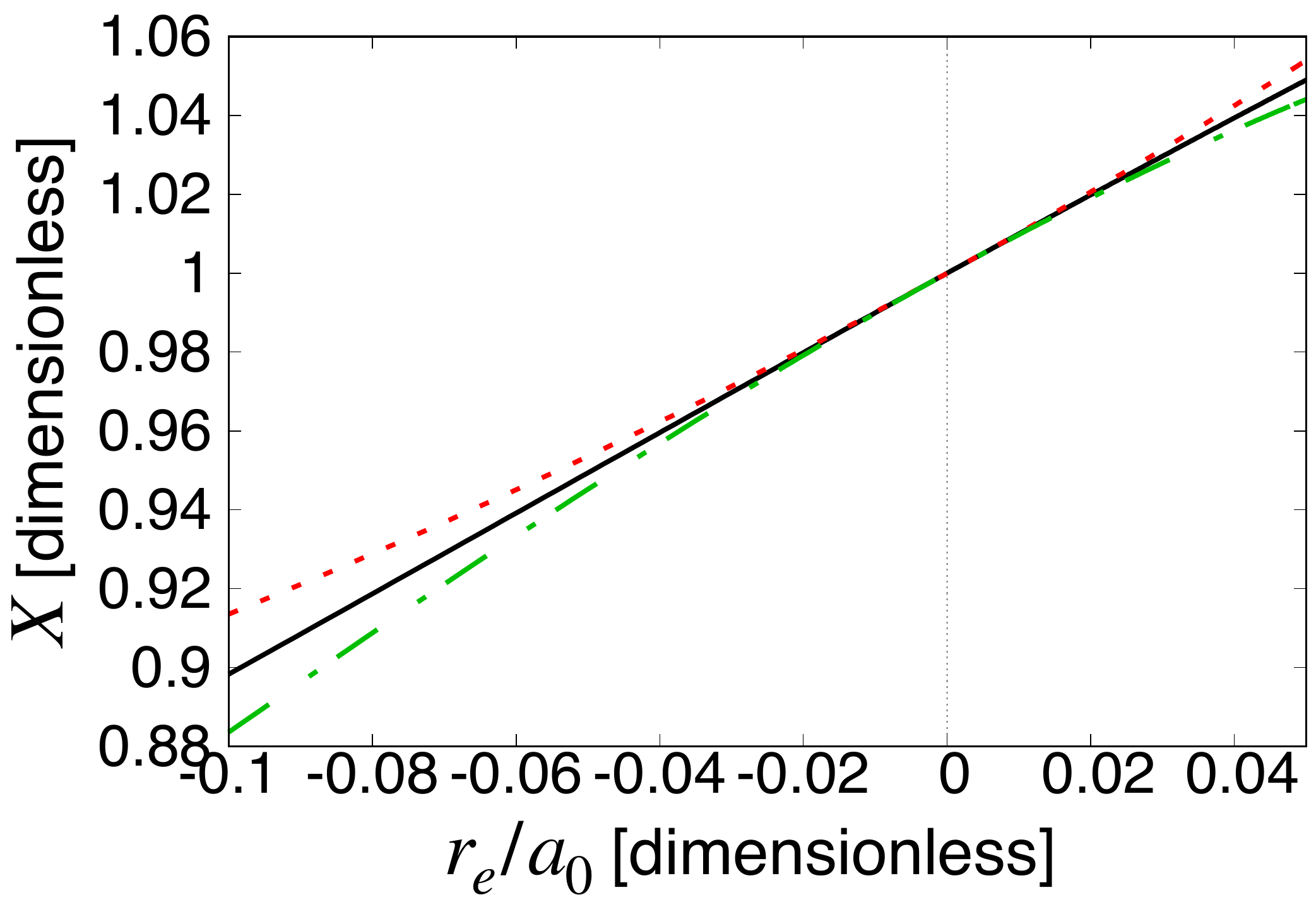}
  \caption{The comparison of the central values of the compositeness $X_{c}$ [Eq.~\eqref{eq:Xc}, solid line], $X_{c}(r_{e},R)$ [Eq.~\eqref{eq:X-re-R-4}, dotted line] and $X_{c}(a_{0},r_{e},R)$ [Eq.~\eqref{eq:X-a0-re-R-4}, dash-dotted line] of the weak-binding relations. \label{fig:Xcs}}
\end{figure}

We discuss the contributions from the higher order terms of the effective range expansion by numerically comparing the improved central value [Eq.~\eqref{eq:X-a0-re-R-4}] with the original one [Eq.~\eqref{eq:Xc}]. We use the effective range model with the finite cutoff ($\tilde{R}_{\rm int}=0.5$) for the calculation. Figure~\ref{fig:Xcs} shows the central values of the compositeness $X_{c}$: the original one [Eq.~\eqref{eq:Xc}, solid line], $X_{c}(r_{e},R)$ [Eq.~\eqref{eq:X-re-R-4}, dotted line], and $X_{c}(a_{0},r_{e},R)$ [Eq.~\eqref{eq:X-a0-re-R-4}, dash-dotted line]. We observe $X_{c}=1$ at $\tilde{r}_{e}=0$ in all cases. Equation~\eqref{eq:Xc} induces $X_{c}=1$ because the condition $\tilde{r}_{e}=0$ corresponds to the zero-range model with $a_{0}=R$. The higher order terms in the effective range expansion also vanish for $\tilde{r}_{e}=0$ because the bare parameter $\tilde{\rho}_{0}$ becomes zero as shown in Fig.~\ref{fig:re-rho0}. Therefore, Eqs.~\eqref{eq:X-re-R-4} and \eqref{eq:X-a0-re-R-4} reduce to Eq.~\eqref{eq:Xc} because the differences arise from the higher order terms, and $X_{c}=1$ holds in all cases.
 We see from Fig.~\ref{fig:Xcs} that the deviations of Eqs.~\eqref{eq:X-re-R-4} and \eqref{eq:X-a0-re-R-4} from Eq.~\eqref{eq:Xc}
arising from the $P_{s}$ term, are of the same order. Improved $X_{c}(a_{0},r_{e},R)$ is closer to $X_{\rm exact}=1$ than the original one in the $\tilde{r}_{e}>0$ region. In contrast, the deviation of $X_{c}(a_{0},r_{e},R)$ from $X_{\rm exact}=1$ is large in the $\tilde{r}_{e}<0$ region. While the contribution of $P_{s}$ enlarges the applicable region for $\tilde{r}_{e}>0$, it decreases for $\tilde{r}_{e}<0$. Therefore, the contribution of the higher order terms in Eq.~\eqref{eq:X-a0-re-R-4} does not always improve the weak-binding relation.
In fact, the differences in Fig.~\ref{fig:Xcs} are much smaller than the width of the uncertainty bands in Fig.~\ref{fig:X_boundary}(d). While the deviation from the original $X_{c}$ [Eq.~\eqref{eq:Xc}] is about 0.015 at $\tilde{r}_{e}\approx -0.1$ in Fig.~\ref{fig:Xcs}, the width of the uncertainty band by $\xi_{\rm int}$ is about 2.  Therefore, the order of the deviation of Eq.~\eqref{eq:X-a0-re-R-4} from Eq.~\eqref{eq:Xc} is negligibly small, and the contribution of the higher order terms of the effective range expansion in Eq.~\eqref{eq:X-a0-re-R-4} does not affect the compositeness estimation. We obtain the same results for $\tilde{R}_{\rm int}\leq 0.5$. 

\section{Application to physical systems}
\label{sec:apply}
We apply the improved weak-binding relation with the range correction~\eqref{eq:redef} to the actual physical systems to investigate their internal structure. Because the weak-binding relation is based on the consequence of the low-energy universality, we can use it for any shallow bound states at arbitrary length scales. Here we apply the weak-binding relation to the hadrons, nuclei, and atomic systems. In Sec.~\ref{subsec:a0-re}, we summarize the observables and the interaction range in the systems to which we apply the weak-binding relation. We then discuss the importance of the range correction in these systems in Sec.~\ref{subsec:X-effectiverangemodel}. Finally, we estimate the compositeness to discuss the internal structure of the states in Sec.~\ref{subsec:estimation-X}.

\subsection{Properties of bound states}
\label{subsec:a0-re}
We summarize the properties of the bound states we consider in Table~\ref{tab:mass}. The deuteron $d$ in the $pn$ scattering is chosen as an example of the experimentally observed hadron. We focus on $X(3872)$, $D^{*}_{s0}(2317)$, and $D_{s1}(2460)$ as the representative candidates for the exotic hadrons~\cite{ParticleDataGroup:2020ssz}. These hadrons are regarded as a bound state of $D^{0}\bar{D}^{*0}$, $DK$, and $D^{*}K$, respectively.\footnote{While the wavefunction of the $D$-meson pair coupled with $X(3872)$ is the linear combination $(D^{0}\bar{D}^{*0}+\bar{D}^{0}D^{*0})/\sqrt{2}$, here we denote it as $D^{0}\bar{D}^{*0}$ for simplicity.} In the recent lattice QCD~\cite{HALQCD:2018qyu,Gongyo:2017fjb}, $N\Omega$ and $\Omega \Omega$ systems are predicted to have a shallow bound state in the ${}^{5}S_{2}$ and ${}^{1}S_{0}$ channel, respectively. We apply the weak-binding relation to those theoretically predicted states. As an example of the hypernuclei, we consider the ${}^{3}_{\Lambda}{\rm H}$ as a bound state of $d$ and $\Lambda$ with a small binding energy. In addition to the hadrons and nuclei, the atomic states can also be considered. The ${}^{4}{\rm He}$ dimer is the shallow bound state of two ${}^{4}{\rm He}$ atoms. 
In Table~\ref{tab:mass}, we show the masses of the two-body channel, $m_{1}$ and $m_{2}$, and the binding energy $B$. The masses and the binding energies of the observed hadrons including $d$ are taken from the Particle Data Group~\cite{ParticleDataGroup:2020ssz}. The masses of $K$, $D$, and $D^{*}$ are averaged over the isospin multiplet.  We adopt the binding energies $B$ of the $N\Omega$ and $\Omega \Omega$ dibaryons by lattice QCD~\cite{HALQCD:2018qyu,Gongyo:2017fjb}. To be consistent with the calculation of $B$, we use the masses in the $N\Omega$ and $\Omega\Omega$ systems given by the same lattice setup~\cite{HALQCD:2018qyu,Gongyo:2017fjb}. The binding energy $B$ of ${}^{3}_{\Lambda}{\rm H}$ is obtained from the emulsion data~\cite{Juric:1973zq}. We refer to Refs.~\cite{Wang:2021xhn} and \cite{Braaten:2004rn} for the mass of the ${}^{4}{\rm He}$ atom and the binding energy of the ${}^{4}{\rm He}$ dimer, respectively.

 \begin{table*}
 \caption{The summary of the properties of the bound states that we consider in this work. We estimate the compositeness of the bound states with respect to the two-body channel of particles 1 and 2 with the masses $m_{1}$ and $m_{2}$. Shown is the binding energy $B$ measured from the threshold of particles 1 and 2, the scattering length $a_{0}$, the effective range $r_{e}$, the radius of the bound state $R=1/\sqrt{2\mu B}$, and the interaction range $R_{\rm int}$. u, mK, and  B.R. stand for the atomic mass unit, millikelvin, and the Bohr radius. \label{tab:mass}}
 \begin{ruledtabular}
  \begin{tabular}{ccccccccccc}
    Bound state& Particle 1 & Particle 2 & $m_{1}$ & $m_{2}$ & $B$ & $a_0$ & $r_e$ & $R$ & $R_{\rm int}$ \\ \hline 
    $d$ & $p$ & $n$ &  938.3 MeV & 939.6 MeV & $2.22$ MeV & 5.42 fm & 1.75 fm & 4.32 fm & $1.43$\ fm\\ 
      $X(3872)$ & $D^{0}$ & $\bar{D}^{*0}$ & 1865 MeV & 2010 MeV & 0.018 MeV &28.5 fm & $-5.34$ fm & 33.4 fm & $1.43$\ fm\\ 
       $D^{*}_{s0}(2317)$ & $D$ & $K$ & 1867 MeV & 495.6 MeV & 44.8 MeV & 1.3 fm & $-$0.1 fm & 1.05 fm & 0.359 fm\\ 
        $D_{s1}(2460)$ & $D^{*}$ & $K$ & 2009 MeV & 495.6 MeV & 45.1 MeV & 1.1 fm & $-$0.2 fm & 1.03 fm & 0.359 fm\\ 
        $N\Omega$ dibaryon & $N$ & $\Omega $ &  955 MeV & 1712 MeV & $1.54$\ MeV & 5.30\ fm&1.26\ fm & 4.54 fm & $0.676$ fm\\ 
        $\Omega \Omega$ dibaryon & $\Omega $ & $\Omega $ & 1712 MeV & 1712 MeV & $1.6$\ MeV  & 4.6\ fm &1.27\ fm& 3.77 fm & $0.949$ fm \\ 
        ${}^{3}_{\Lambda}{\rm H}$ & $d$ & $\Lambda$ &1876 MeV & 1116 MeV & $0.13$\ MeV & 16.8\ fm&2.3\ fm& 14.6 fm& $4.32$ fm\\ 
        ${}^{4}{\rm He}$ dimer &${}^{4}{\rm He}$ &${}^{4}{\rm He}$&4.003 u&4.003 u & $1.30$\ mK & 189\ B.R.&13.8\ B.R.& 182.2 B.R. & $10.2$ B.R. \\ 
  \end{tabular}
   \end{ruledtabular}
 \end{table*} 

In general, the weak-binding relation cannot be applied to the systems with long-range interaction such as the Coulomb force. Because $p$, $\Omega$, and $d$ are the charged particles in Table~\ref{tab:mass}, the Coulomb force can be active in the $N\Omega$ and $\Omega \Omega$ channels.\footnote{Because $D^{*}_{s0}(2317)$ and $D_{s1}(2460)$ have charge $\pm 1$, one of the particles in the two-body system is neutral. Therefore, the Coulomb force does not operate in these systems.} When we choose the neutron $\Omega$ scattering in the isospin doublet of the $N\Omega$ system, we can avoid the existence of the Coulomb interaction. However, the Coulomb interaction is unavoidable in the physical $\Omega \Omega$ system because both particles are charged. Here we take the lattice results without the Coulomb interaction in Ref.~\cite{Gongyo:2017fjb} to apply the weak-binding relation. We note that the properties of the physical $p\Omega$, and $\Omega \Omega$ dibaryons should be calculated with the Coulomb interaction. 

Strictly speaking, the weak-binding relation~\eqref{eq:wbr_s} is applicable only to the stable bound states. Although $X(3872)$, $D^{*}_{s0}(2317)$, $D_{s1}(2460)$, and the $N\Omega$ dibaryon are unstable, those are expected to have a narrow decay width compared with that of typical unstable hadrons ($\approx 100$ MeV). In fact, the widths of $X(3872)$, $D^{*}_{s0}(2317)$, and $D_{s1}(2460)$ are experimentally determined as $1.19\pm 0.21$ MeV, $<3.8$ MeV and $<3.5$ MeV, respectively~\cite{ParticleDataGroup:2020ssz}, and that of the $N\Omega$ dibaryon is calculated as $\approx1$ MeV by the meson exchange model~\cite{Sekihara:2018tsb}. The narrow decay widths of $D^{*}_{s0}(2317)\to D_{s}\pi$ and $D_{s1}(2460)\to D^{*}_{s}\pi$ can be  understood from the isospin breaking nature of the decays. The decay of the $N\Omega$ dibaryon is suppressed because of the large angular momentum $\ell=2$. Therefore, we apply the weak-binding relation to those states by neglecting the effects of the tiny decay widths.

We also show the scattering length $a_{0}$, the effective range $r_{e}$, the radius of the bound state $R$, and the interaction range $R_{\rm int}$ in Table~\ref{tab:mass}. The radius $R=1/\sqrt{2\mu B}$ is calculated with $\mu=m_{1}m_{2}/(m_{1}+m_{2})$ and $B$. For the deuteron, $a_{0}$ and $r_{e}$ are taken from the results of the CD-Bonn potential~\cite{Machleidt:2000ge}. For $D^{*}_{s0}(2317)$ and $D_{s1}(2460)$, $a_{0}$ and $r_{e}$ are obtained from the lattice QCD~\cite{MartinezTorres:2014kpc}. For the $N\Omega$ and $\Omega \Omega$ dibaryons, the observables are calculated with the lattice QCD potential in Refs.~\cite{HALQCD:2018qyu,Gongyo:2017fjb}. For ${}^{3}_{\Lambda}{\rm H}$, $a_{0}$ and $r_{e}$ are obtained in the framework of the effective field theory in Ref.~\cite{Hammer:2001ng}. For the ${}^{4}{\rm He}$-${}^{4}{\rm He}$ interaction, the realistic potential (LM2M2 potential) is available~\cite{LM2M2}. Here we use the values of $a_{0}$ and $r_{e}$ calculated with this potential in Ref.~\cite{Kievsky:2012ss}. 

The interaction range $R_{\rm int}$ is not an observable and we determine it with the theoretical consideration of the microscopic structure of the interaction. The interaction range $R_{\rm int}$ for $d$ and $X(3872)$ is given by the Compton wavelength of the pion because the pion exchange is possible in the $pn$ and $D^{0}\bar{D}^{*0}$ scatterings. Similarly, the range of the $DK$ and $D^{*}K$ interactions for $D^{*}_{s0}(2317)$ and $D_{s1}(2460)$ is estimated by the $\sigma$-meson exchange with $m_{\sigma}\approx 550$ MeV. The interaction range of the $N\Omega$ system is estimated by the longest length scale $R_{\rm int}\approx 1/2m_{\pi}^{\rm lat}$ in the HAL QCD potential $V_{\rm fit}(r)=b_{1}\exp[-b_{2}r^{2}]+b_{3}(1-\exp[-b_{4}r^{2}])^{n}(e^{-m_{\pi}^{\rm lat}r}/r)^{2}$ with $m_{\pi}^{\rm lat}=146$ MeV in the lattice QCD calculation~\cite{HALQCD:2018qyu}. For the $\Omega \Omega$ interaction, we obtain $R_{\rm int}$ by the largest length scale $R_{\rm int}\approx d_{3}= 0.949$ fm in the lattice QCD potential $V_{\rm fit}(r)=\sum_{j=1,2,3}c_{j}\exp[-(r/d_{j})^{2}]$~\cite{Gongyo:2017fjb}. Because of the short range nature of the $\Lambda N$ interaction, it is expected that $d$ interacts with $\Lambda$ when $\Lambda$ overlaps with the density distribution of $d$. Therefore, the radius of $d$ is regarded as $R_{\rm int}$ of the $d$-$\Lambda$ interaction. The interaction range of ${}^{4}{\rm He}$ dimer is estimated by the van der Waals length $R_{\rm int}\approx l_{\rm vdW}=(m{\rm C}_{6}/\hbar^{2})^{1/4}$, where C${}_{6}$ is obtained by the potential at $r\to\infty$ [$V(r)\to {\rm C}_{6}/r_{6}$] in Ref.~\cite{Yan:1996}. In Table~\ref{tab:mass}, we find that in all cases $a_{0}$ is much larger than $|r_{e}|$ and $R_{\rm int}$. This fact can be clearly seen by the dimensionless quantities $\tilde{r}_{e}=r_{e}/a_{0}$ and $\tilde{R}_{\rm int}=R_{\rm int}/a_{0}$ whose magnitudes are much smaller than unity in Table~\ref{tab:applicable-systems-parameter}. By introducing these dimensionless quantities, the systems with the different length scales (hadron and atomic systems) can be treated on the same footing.

 \begin{table}
 \caption{The dimensionless effective range $\tilde{r}_{e}=r_{e}/a_{0}$ and the dimensionless interaction range $\tilde{R}_{\rm int}=R_{\rm int}/a_{0}$ of the two-body systems coupled with the bound state. \label{tab:applicable-systems-parameter}}
 \begin{ruledtabular}
  \begin{tabular}{ccc} 
   Bound state&$\tilde{r}_{e}$ & $\tilde{R}_{\rm int}$ \\ \hline
    $d$ & 0.323 & 0.264\\ 
     $X(3872)$ & $-0.187$ & 0.0501\\ 
     $D^{*}_{s0}(2317)$ & $-0.077$ & 0.28\\ 
     $D_{s1}(2460)$ & $-0.18$ & 0.33\\ 
        $N\Omega$ dibaryon& 0.238 & 0.128\\ 
        $\Omega \Omega$ dibaryon& 0.276 & 0.206\\  
        ${}^{3}_{\Lambda}{\rm H}$ & 0.137 & 0.257 \\ 
         ${}^{4}{\rm He}$ dimer & 0.0730 & 0.0540 \\ 
  \end{tabular}
 \end{ruledtabular}
 \end{table}

\subsection{Comparison with applicable region of effective range model}
\label{subsec:X-effectiverangemodel}

Before the estimation of the compositeness, we discuss the significance of the range correction to the weak-binding relation by using the actual systems in Table~\ref{tab:mass}. For this purpose, we use the applicable regions of the weak-binding relations in the effective range model (Fig.~\ref{fig:parameter-regions}). 

\begin{figure}[t]
\centering
\includegraphics[width=0.45\textwidth]{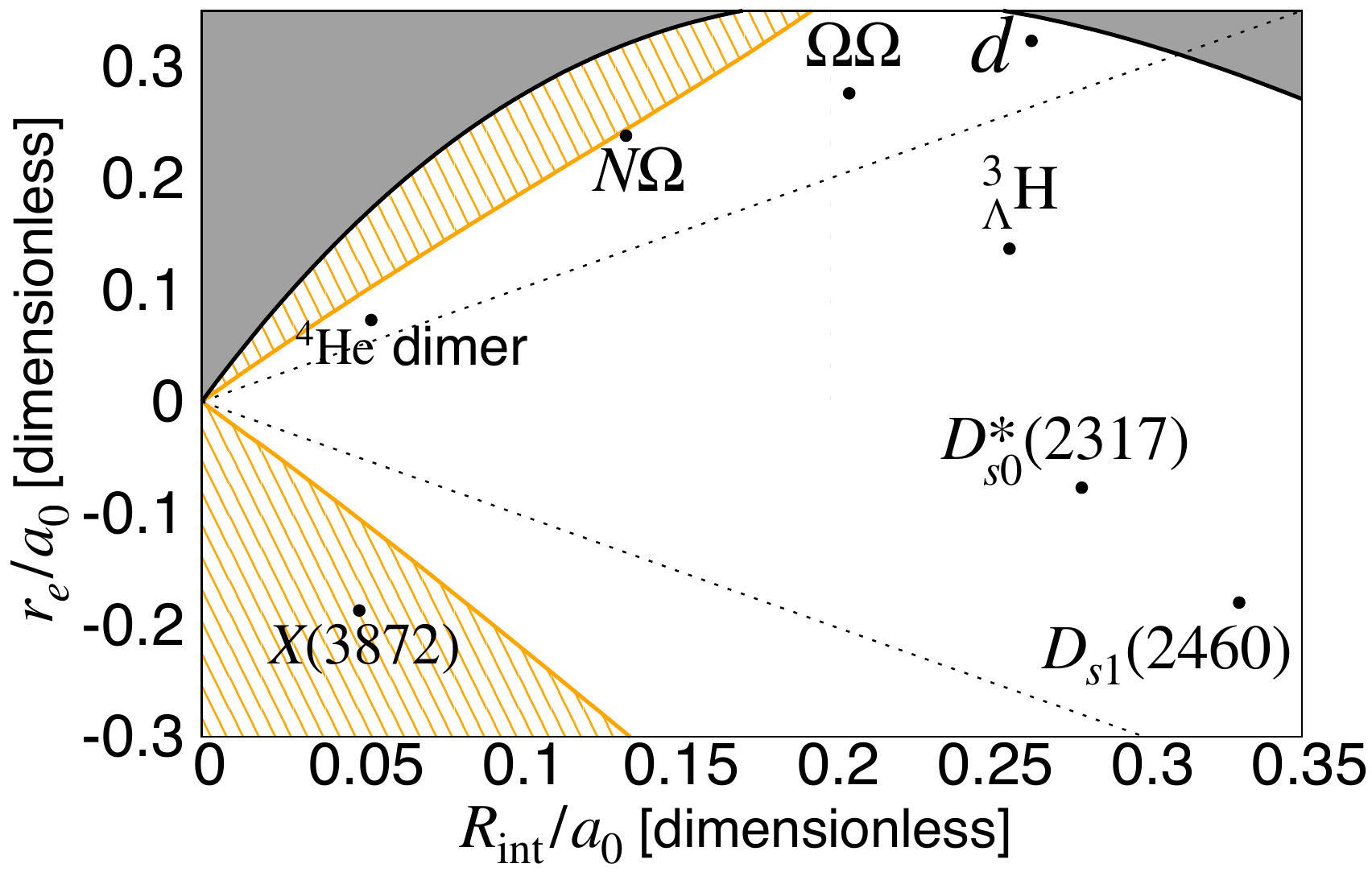}
\caption{Comparison of the bound states with the applicable regions in the effective range model in the $\tilde{R}_{\rm int}$-$\tilde{r}_{e}$ plane. The legends are the same as Fig.~\ref{fig:parameter-regions}.}
\label{fig:applicable-region-hadron}
\end{figure}

In Fig.~\ref{fig:applicable-region-hadron}, we compare the parameters listed in Table~\ref{tab:applicable-systems-parameter} with the applicable regions in the effective range model in the $\tilde{R}_{\rm int}$-$\tilde{r}_{e}$ plane. The dotted lines stand for $\tilde{r}_{e}=\pm \tilde{R}_{\rm int}$. For the previous weak-binding relation~\eqref{eq:wbr_s}, the correction terms $\mathcal{O}(R_{\rm typ}/R)$ are always estimated by $R_{\rm typ}=R_{\rm int}$. However, for the improved weak-binding relation with the range correction~\eqref{eq:Reff}, we adopt the largest length scale among $R_{\rm int}$ and $R_{\rm eff}$ as $R_{\rm typ}$. Hence, we find from Fig.~\ref{fig:applicable-region-hadron} that $R_{\rm typ}=R_{\rm int}$ for $D^{*}_{s0}(2317)$, $D_{s1}(2460)$, and ${}^{3}_{\Lambda}{\rm H}$ but $R_{\rm typ}=R_{\rm eff}$ for all other states. This indicates that the range correction plays an important role for some physical systems. In fact, $X(3872)$ is in the region where only the improved weak-binding relation with the range correction is applicable. Therefore, the previous relation may fail to estimate the compositeness of $X(3872)$, and we should use the improved relation for the reasonable estimation. In addition, because the $N\Omega$ dibaryon and ${}^{4}$He dimer lie close to the applicable boundary of the previous weak-binding relation, it is also expected that the range correction is quantitatively important for these states. We should, however, keep in mind that the applicable region in Fig.~\ref{fig:applicable-region-hadron} is the result of the specific model (the effective range model) and the applicable boundaries are model dependent. 

\begin{figure}[t]
\centering
\includegraphics[width=0.45\textwidth]{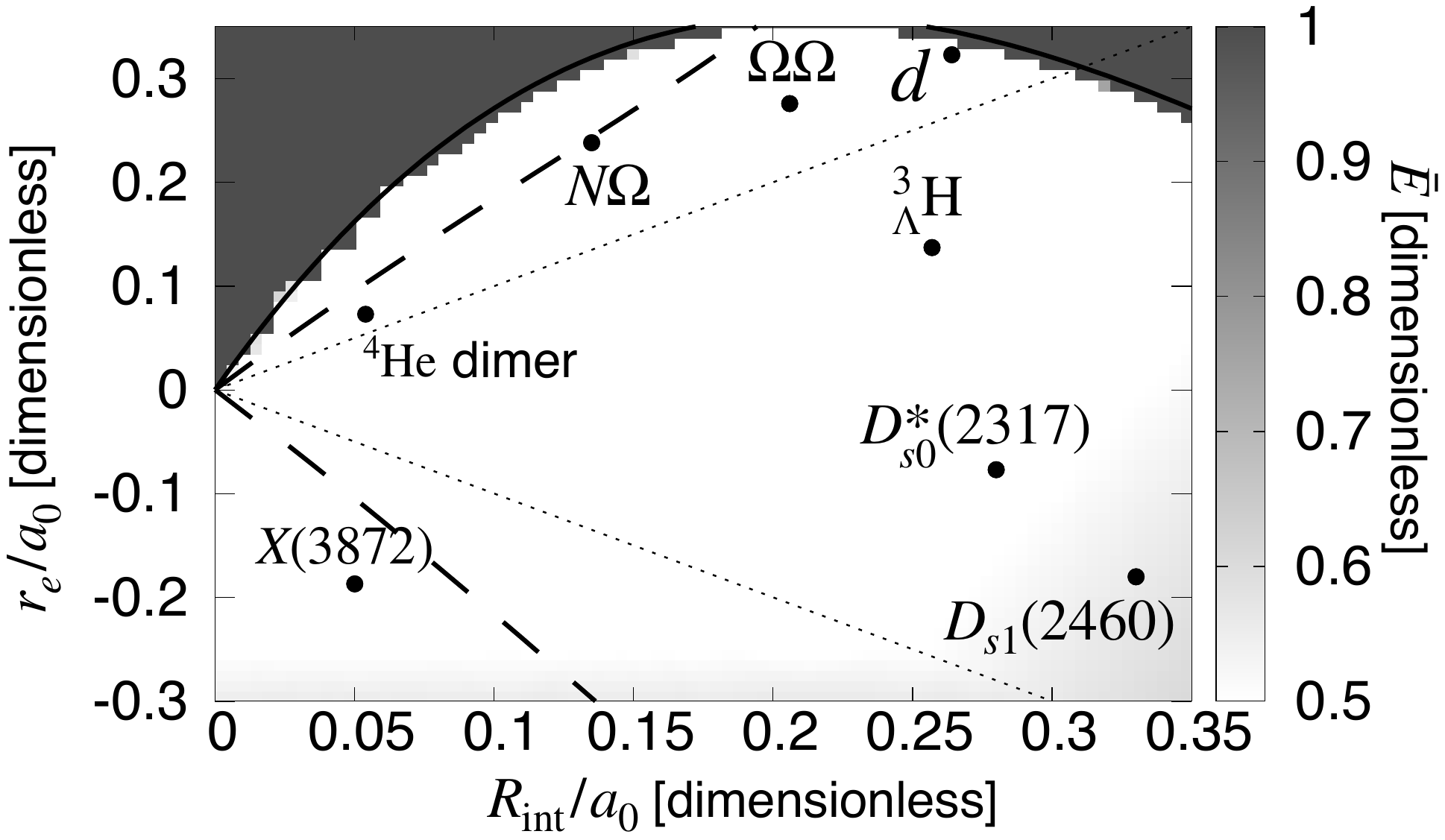}
\caption{
Comparison of the bound states with the distribution of the magnitude of the uncertainty $\bar{E}$ in the effective range model in the $\tilde{R}_{\rm int}$-$\tilde{r}_{e}$ plane. The legends are the same as Fig.~\ref{tab:Ebar-r-Lambda}.}
\label{fig:barE-hadrons}
\end{figure}

For the discussion of the meaningful estimation, we plot the parameters in Table~\ref{tab:applicable-systems-parameter} in comparison with the magnitude of the uncertainty $\bar{E}$ in $\tilde{R}_{\rm int}$-$\tilde{r}_{e}$ plane in Fig.~\ref{fig:barE-hadrons}. Because all the states are contained in the region $\bar{E}\lesssim0.5$, we expect that meaningful estimations of the compositeness are possible for these states. 

\subsection{Estimation of compositeness}
\label{subsec:estimation-X}
We now estimate the compositeness $X$ of the bound states listed in Table~\ref{tab:mass}. We summarize the estimated uncertainties $\xi_{\rm eff}=|r_{e}|/R$ and $\xi_{\rm int}=R_{\rm int}/R$ in Table~\ref{tab:estimated-X}. Here, we set $R_{\rm eff}=|r_{e}|$ assuming that the coefficients of the higher order terms in the effective range expansion are of natural size. We then show the estimated compositeness with the uncertainty band with $\xi_{\rm eff}$ [$X(\xi_{\rm eff})$] and $\xi_{\rm int}$ [$X(\xi_{\rm int})$] in Table~\ref{tab:estimated-X}. In the last column we also show $R_{\rm typ}$ in the improved weak-binding relation~\eqref{eq:redef}.

We can see that the central values of the compositeness $X_{c}$ are larger than unity except for $X(3872)$ in Table~\ref{tab:estimated-X}. This is because the radius $R$ is smaller than the scattering length $a_{0}$ in these states. As we discussed in Sec.~\ref{sec:rangecorrectionsdiscussiton}, $X_{c}$ is larger than unity for $a_{0}>R$. 
The relation between $a_{0}$ and $R$ is also approximately determined by the sign of $r_{e}$. Neglecting the $\mathcal{O}(k^{4})$ terms in the effective range expansion, we obtain Eq.~\eqref{eq:relations}: 
\begin{align}
a_{0}&=R\frac{1}{-r_{e}/(2R)+1}.
\label{eq:a-R-re}
\end{align}
Because $R>0$, we obtain $a_{0}>R$ for positive $r_{e}>0$, and $a_{0}<R$ for negative $r_{e}<0$ from this equation. In fact, in Table~\ref{tab:mass}, these relations are satisfied except for $D^{*}_{s0}(2317)$ and $D_{s1}(2460)$ with the small effective range. In summary, the central value of the compositeness is  larger than unity for $a_{0}>R$, which is expected to be realized with positive $r_{e}>0$ when relation~\eqref{eq:a-R-re} approximately holds.

One may wonder that the central value $X_{c}>1$ contradicts with the definition of the compositeness $0\leq X\leq1$. In fact, this problem for the deuteron partly motivates the works in Refs.~\cite{Li:2021cue,Song:2022yvz,Albaladejo:2022sux}. From our viewpoint, this problem can be avoided by considering the uncertainty $\xi$ as in Eq.~\eqref{eq:Xfinal} as discussed below.

Focusing on the $N\Omega$ dibaryon, we find that the lower limit of the compositeness estimated by $\xi_{\rm int}$ is larger than unity [$X_{l}(\xi_{\rm int})=1.04$] from Table~\ref{tab:estimated-X}. Hence, the exact value of the compositeness of the $N\Omega$ dibaryon is not contained in the uncertainty band of $X(\xi_{\rm int})$, and we cannot perform the meaningful estimation of the compositeness of the $N\Omega$ dibaryon with the previous weak-binding relation ($R_{\rm typ}=R_{\rm int}$). In fact, we have seen that the $N\Omega$ dibaryon exists near the boundary of the applicable region of the previous weak-binding relation in the effective range model as shown in Fig.~\ref{fig:applicable-region-hadron}. 

 \begin{table}
 \caption{The uncertainties $\xi_{\rm eff}$, $\xi_{\rm int}$, 
  the estimated compositeness $X$, and the length scale $R_{\rm typ}$ in the improved weak-binding relation. $X(\xi_{\rm eff})$ [$X(\xi_{\rm int})$] 
 stands for $X$ estimated with $\xi_{\rm eff}$ ($\xi_{\rm int}$).\label{tab:estimated-X}}
 \begin{ruledtabular}
  \begin{tabular}{cccccc} 
    Bound state& $\xi_{\rm eff}$ & $\xi_{\rm int}$ & $X(\xi_{\rm eff})$ & $X(\xi_{\rm int})$ & $R_{\rm typ}$\\ \hline 
    $d$ & 0.405 & 0.331 & $1.68^{+3.18}_{-0.943}$ & $1.68^{+2.14}_{-0.824}$ & $R_{\rm eff}$\\ 
      $X(3872)$ & 0.160 & 0.0428 & $0.743^{+0.282}_{-0.213}$ & $0.743^{+0.0675}_{-0.0626}$ & $R_{\rm eff}$\\ 
       $D^{*}_{s0}(2317)$ & 0.0949 & 0.341 & $1.61^{+0.369}_{-0.288}$ & $1.61^{+2.09}_{-0.804}$ & $R_{\rm int}$\\ 
        $D_{s1}(2460)$ & 0.192 & 0.345 & $1.12^{+0.540}_{-0.358}$ & $1.12^{+1.22}_{-0.566}$ & $R_{\rm int}$\\ 
        $N\Omega$ dibaryon & 0.277 & 0.149 & $1.40^{+1.20}_{-0.600}$ & $1.40^{+0.523}_{-0.364}$ & $R_{\rm eff}$\\ 
        $\Omega \Omega$ dibaryon & 0.337 & 0.252 & $1.56^{+1.95}_{-0.773}$ & $1.56^{+1.22}_{-0.626}$ & $R_{\rm eff}$\\  
        ${}^{3}_{\Lambda}{\rm H}$ & 0.157 & 0.295 & $1.35^{+0.532}_{-0.366}$ & $1.35^{+1.25}_{-0.605}$ & $R_{\rm int}$ \\ 
        ${}^{4}{\rm He}$ dimer & 0.0757 & 0.0560 & $1.08^{+0.177}_{-0.152}$ & $1.08^{+0.128}_{-0.114}$ & $R_{\rm eff}$\\
  \end{tabular}
 \end{ruledtabular}
 \end{table}

In  the improved weak-binding relation with Eq.~\eqref{eq:Reff}, we calculate compositeness with the uncertainty band as $X(\xi_{\rm eff})$ [$X(\xi_{\rm int})$] for $R_{\rm typ}=R_{\rm eff}$ ($R_{\rm typ}=R_{\rm int}$). From the last column, we see that $X(\xi_{\rm int})$ is adopted for the states $D^{*}_{s0}(2317)$, $D_{s1}(2460)$ and ${}^{3}_{\Lambda}$H, and $X(\xi_{\rm eff})$ for other states. By taking the region consistent with the definition $0\leq X\leq 1$ in Eq.~\eqref{eq:Xfinal}, we finally determine the compositeness $X$ as shown in Table~\ref{tab:compositeness}.

 These results ($0.5\leq X\leq 1$) indicate that the composite component gives the largest fraction in the wavefunction for all states. In particular, the ${}^{4}{\rm He}$ dimer is an almost purely composite state with a small fraction of the other components ($\lesssim 7\ \%$). However, the compositeness of $X(3872)$ and $D_{s1}(2460)$ can be as low as $\approx$0.5, which is the boundary of the composite dominance. Therefore, it is expected that the other components would play a substantial role in these states. We find that Eq.~\eqref{eq:Xfinal} gives a reasonable estimation of the compositeness of the deuteron $0.74\leq X\leq 1$, indicating its composite nature. The compositeness of the $N\Omega$ dibaryon is also meaningfully estimated thanks to the range correction~\eqref{eq:redef}.

 \begin{table}
 \caption{The compositeness $X$ consistent with the definition~\eqref{eq:Xfinal} estimated by the improved weak-binding relation.\label{tab:compositeness}}
 \begin{ruledtabular}
  \begin{tabular}{cc}  
    Bound state& Compositeness$X$ \\ \hline 
    $d$ & $0.74\leq X\leq 1$  \\
      $X(3872)$ & $0.53\leq X\leq 1$  \\
	$D^{*}_{s0}(2317)$ & $0.81\leq X\leq 1$  \\ 
       $D_{s1}(2460)$ & $0.55\leq X\leq 1$  \\  
       $N\Omega $ dibaryon& $0.80\leq X\leq 1$  \\ 
        $\Omega \Omega $ dibaryon& $0.79\leq X\leq 1$  \\  
        ${}^{3}_{\Lambda}{\rm H}$ & $0.74\leq X\leq 1$   \\ 
        ${}^{4}{\rm He}$ dimer& $0.93\leq X\leq 1$  \\
  \end{tabular}
 \end{ruledtabular}
 \end{table}

We compare our results with that of the previous works focusing on the deuteron $d$, $X(3872)$, $D^{*}_{s0}(2317)$ and $D_{s1}(2460)$. The pioneering work for $d$ by Weinberg~\cite{Weinberg:1965zz} concluded that $d$ was the composite state while the quantitative determination of the compositeness was not given. In Ref.~\cite{Kamiya:2017hni}, the compositeness of $d$ was quantitatively calculated as $1.68^{+2.15}_{-0.83}$ with the uncertainty estimation from the correction terms in the previous weak-binding relation. This result corresponds to $X(\xi_{\rm int})$ in Table~\ref{tab:estimated-X}. By taking the range correction into account appropriately, we find the uncertainty band of the compositeness of $d$, shown as $X(\xi_{\rm eff})$ in Table~\ref{tab:estimated-X}, is larger than that of Ref.~\cite{Kamiya:2017hni}. In recent works~\cite{Li:2021cue,Song:2022yvz,Albaladejo:2022sux}, the deuteron was found to be composite dominant. In particular, Ref.~\cite{Albaladejo:2022sux} concluded that values of $X$ smaller than $\approx 0.7$ were very implausible. Those results are qualitatively consistent with our estimation $0.74\leq X\leq 1$. 

The compositeness of $D^{*}_{s0}(2317)$ and $D_{s1}(2460)$ is discussed in Refs.~\cite{Song:2022yvz,Albaladejo:2022sux,MartinezTorres:2014kpc}. For $D^{*}_{s0}(2317)$, Refs.~\cite{Song:2022yvz}, \cite{Albaladejo:2022sux}, and \cite{MartinezTorres:2014kpc} found that $X>0.6$, $X>0.5$, and $X\approx 0.72$, respectively. For $D_{s1}(2460)$, Ref.~\cite{Song:2022yvz} and  Ref.~\cite{MartinezTorres:2014kpc} found that $0.4<X<0.7$ and  $X\approx 0.57$, respectively. Our results [$0.81\leq X\leq 1$ for $D^{*}_{s0}(2317)$ and $0.55\leq X\leq1$ for $D_{s1}(2460)$] are similar to the previous works; $D^{*}_{s0}(2317)$ is relatively composite dominant, and $D_{s1}(2460)$ can contain an appreciable amount of the noncomposite components. The quantitative difference of the results of $D_{s1}(2460)$ may be attributed to the large uncertainty $\xi_{\rm int}=0.345$ which indicates that the binding energy of $D_{s1}(2460)$ is not sufficiently small. We note that the inputs $a_{0}$ and $r_{e}$ for the charmed-strange mesons still have a large uncertainty, for instance $a_{0}(KD)=+1.3\pm 0.5\pm 0.1$ fm and $r_{e}(KD)=-0.1 \pm 0.3\pm 0.1$ fm~\cite{MartinezTorres:2014kpc}.

The structure of $X(3872)$ was studied in the hybrid model of $c\bar{c}$ and hadronic molecules~\cite{Takizawa:2012hy}. Assuming the wavefunction of $X(3872)$ as 
\begin{align}
\ket{X(3872)}&=c_{1}\ket{c\bar{c}}+c_{2}\ket{D^{0}\bar{D}^{*0}}+c_{3}\ket{D^{+}D^{*-}},
\end{align}
they determined the coefficients $c_{i}$ from the comparison with the experiments which lead to $-0.947\leq c_{2}\leq -0.871$. Because the $D^{0}\bar{D}^{*0}$ compositeness of $X(3872)$ corresponds to $X=|c_{2}|^{2}$, this result is interpreted as $0.759\leq X\leq0.897$. Our model-independent result  $0.53\leq X\leq1$ contains that of the model calculation~\cite{Takizawa:2012hy}, as expected.

\section{Summary}
\label{sec:sum}

In this work, we have discussed the range correction to the weak-binding relation for the systems with a large effective range. We introduce an effective field theory to deal with the bound states in various models. Based on the effective range model in the zero range limit, we show the necessity of the range correction in the weak-binding relation. A prescription of the range correction is presented as the redefinition of $R_{\rm typ}$ in the correction terms as the maximum length scale among the interaction range $R_{\rm int}$ and the length scale in the effective range expansion $R_{\rm eff}$. This range correction results in the modification of the uncertainty estimation of the compositeness, which should be performed in conjunction with the definition of the compositeness.

The applicability of the weak-binding relations has been studied numerically with the effective range model ($X=1$) and the resonance model ($X<1$). In both cases, we show that the range correction improves the weak-binding relation with the larger applicable region than the previous one. We have also studied the precision of the estimation of $X$ to calculate the magnitude of the uncertainty $\bar{E}$. 

Finally, we study the compositeness of the actual hadrons, nuclei and atomic states by the weak-binding relation with the range correction. All of states we discuss [the deuteron, $X(3872)$, $D^{*}_{s0}(2317)$, $D_{s1}(2460)$, $N\Omega$ dibaryon, $\Omega\Omega$ dibaryon, ${}^{3}_{\Lambda}{\rm H}$, and ${}^{4}{\rm He}$ dimer] are contained in the region where the meaningful estimation can be performed ($\bar{E}\lesssim 0.5$). Our final results in Table~\ref{tab:compositeness} show that these are basically composite dominant states. In particular, the deuteron compositeness is reasonably estimated as $0.74\leq X\leq 1$. The importance of the range correction is found for the $N\Omega $ dibaryon and $X(3872)$ whose compositeness is not properly estimated by the previous weak-binding relation.

For the future prospects, it is important to discuss the range correction to the weak-binding relation for the unstable states because most of the exotic hadron candidates are resonances with a finite decay width. To this end, it is needed to establish the procedures of the numerical calculation for the unstable states with complex quantities. The study of the compositeness of $T_{cc}$ is also an interesting subject, because the magnitude of the effective range is expected to be large~\cite{LHCb:2021auc}. While the current determination of $r_{e}$ still suffers from the large uncertainty, we expect  that the compositeness of $T_{cc}$ can be determined by the weak-binding relation when the uncertainty of $r_{e}$ reduces in future experiments.

\begin{acknowledgments}
The authors thank Yusuke Nishida for the useful discussion on the weak-binding relation and the zero-range interactions during the ECT* workshop in 2019, which partly motivated this work. This work has been supported in part by the Grants-in-Aid for Scientific Research from JSPS (Grants
No. JP22K03637, 
No. JP19H05150, 
No. JP18H05402, and
No. JP16K17694). 
This work was supported by JST, the establishment of university
fellowships towards the creation of science technology innovation, Grant No. JPMJFS2139. 
\end{acknowledgments}

\bibliography{ref}

\end{document}